\documentclass[preprint]{aastex}
\usepackage{subfig}
\usepackage[dvips]{color}

\slugcomment{Draft 20101001 Accepted by ApJ}

\begin{document}

\title{High Velocity Molecular Outflows In Massive Cluster Forming Region G10.6-0.4}

\author{Hauyu Baobab Liu\altaffilmark{1,2,3}}
\affil{Harvard-Smithsonian Center for Astrophysics, 60 Garden Street, Cambridge, MA 02138}
\email{hlu@cfa.havard.edu}

\author{Paul T. P. Ho\altaffilmark{1,2}}
\affil{Academia Sinica Institute of Astronomy and Astrophysics, \\P.O. Box 23-141, Taipei, 106 Taiwan}
\email{pho@asiaa.sinica.edu.tw}

\author{Qizhou Zhang\altaffilmark{2}}
\affil{Harvard-Smithsonian Center for Astrophysics, 60 Garden Street, Cambridge, MA 02138}
\email{qzhang@cfa.harvard.edu}

\altaffiltext{1}{Academia Sinica Institute of Astronomy and Astrophysics}
\altaffiltext{2}{Harvard-Smithsonian Center for Astrophysics}
\altaffiltext{3}{Department of Physics, National Taiwan University}

\begin{abstract}
We report the arcsecond resolution SMA observations of the $^{12}$CO (2-1) transition in the massive cluster forming region G10.6-0.4.
In these observations, the high velocity $^{12}$CO emission is resolved into individual outflow systems, which have a typical size scale of a few arcseconds.
These molecular outflows are energetic, and are interacting with the ambient molecular gas.
By inspecting the shock signatures traced by CH$_{3}$OH, SiO, and HCN emissions, we suggest that abundant star formation activities are distributed over the entire 0.5 pc scale dense molecular envelope.
The star formation efficiency over one global free--fall timescale (of the 0.5 pc molecular envelope, $\sim$10$^{5}$ years) is about a few percent. 
The total energy feedback of these high velocity outflows is higher than 10$^{47}$ erg, which is comparable to the total kinetic energy in the rotational motion of the dense molecular envelope.
From order--of--magnitude estimations, we suggest that the energy injected from the protostellar outflows is capable of balancing the turbulent energy dissipation.
No high velocity bipolar molecular outflow associated with the central OB cluster is directly detected, which can be due to the photo--ionization. 
\end{abstract}


\section{Introduction}
Theoretical studies suggest that high velocity molecular outflows play important roles in massive cluster forming regions.
Massive molecular outflows associated with the accretion of the OB stars can create cavities with low molecular densities.
The efficient photon leakage through the cavities potentially reduces the radiation pressure exerted on the molecular gas by a factor of 10 (Krumholz 2005), and therefore enhances the stellar accretion rate.
However, if those massive molecular outflows emanated from OB stars are oriented in the bipolar direction which has lower molecular gas density, it is unclear how efficient the outflow momentum and energy can propagate and feedback into the majority of dense molecular gas. 
From the numerical hydrodynamical simulations, Li \& Nakamura (2006) and Nakamura \& Li (2007) suggested that in the parsec scale molecular cloud, the initial turbulence dissipates quickly, which leads to local contraction and star formations (see also Carroll et al. 2009). 
The turbulence induced by the outflows from the intermediate mass or solar mass (proto--)stars over the entire parsec scale cloud, however, quickly replenishes the initial turbulence to support the cloud.
The contraction of the cloud is therefore self--regulated with a low star formation efficiency.
In such scenarios, the molecular cloud can be maintained in a quasi--equilibrium state, which may be conducive to the formation of a dense molecular core in its center. 
Massive stars then are formed in the dense molecular core, and create bright ultracompact (UC) H\textsc{ii} regions via the stellar ionization.

Observationally, extensive single dish surveys of the $^{12}$CO emissions found high detection rates of massive molecular outflows around the massive cluster forming regions.
In addition, the ubiquitously seen bipolar profiles of the massive molecular outflows (Shepherd \& Churchwell 1996 b; Beuther et al. 2002; Zhang et al. 2005; Wu et al. 2005; L{\'o}pez-Sepulcre et al. 2009) suggest that the magnetically regulated accretion similar to that in low--mass star formation plays a role in OB star formation.
Besides the massive bipolar outflows, however, owing to the great complexity of the OB cluster forming regions, identifications of outflows from individual intermediate mass or solar mass stars have been difficult, and the improvements rely on detailed analysis of interferometric data.

G10.6-0.4 is a well studied massive cluster forming region at a 6 kpc distance (Caswell et al. 1975; Ho \& Haschick 1981; Ho et al. 1983; Ho et al. 1986;  Keto et al. 1987;  Guilloteau et al. 1988; Keto et al. 1988;  Ho et al. 1994; Keto 1990; Omodaka et al. 1992; Keto 2002; Sollins et al. 2005; Sollins \& Ho 2005;  Keto \& Wood 2006; Keto et al. 2008; Klaassen et al. 2009).
In this region, a high bolometric luminosity (9.2$\times$10$^{5}$ L$_{\odot}$) and bright free-free continuum emission ($\ge$2.4 Jy within a 0.05 pc radius, at 1.3 cm band) were detected, suggesting that a cluster of O-type stars  (O6.5--B0; Ho and Haschick 1981) has formed.
Our previous SMA observations in CH$_{3}$OH J=5 transitions trace a massive rotating envelope  ($\sim$1000 M$_{\odot}$ in the central 0.3 pc $\times$ 0.1 pc region) which shows a biconical cavity (Liu et al. 2010 b).
The biconical cavity may be initially created by powerful (proto--)stellar outflows, and clues may be provided by sensitive molecular line observations to detect the high velocity emissions.

The VLA/eVLA observations of the centimeter free-free continuum emission and the CS (1-0) emission suggest that the biconical cavity is undergoing an expansional motion with the velocity of a few kms$^{-1}$, which may be powered by the ionized gas pressure. 
These results suggest that the pressure of the ionized gas is an important feedback mechanism in the initially low density region.
In the plane of rotation, the dense gas is marginally centrifugally supported (Liu et al. 2010 a), and its global dynamics is not yet severely disturbed by the radiation pressure and the pressure of the ionized gas.
In the central 0.1 pc region, a fast rotating hot toroid encircling an OB cluster is detected (Keto, Ho, \& Haschick 1988; Sollins \& Ho 2005; Liu et al. 2010), which is consistent with dense core formation via a global contraction of a massive envelope.
The overall geometry and the dynamics qualitatively resemble the standard magnetically--regulated core collapse in the low mass star forming region.

To improve the understanding of how the envelope is shaped by the (proto--)stellar outflows, and how the global contraction is regulated, we carried out the interferometric observations of the $^{12}$CO (2-1) line in G10.6-0.4.
We estimate the mass, energy, and momentum budgets in the detected outflow systems.
In addition, by comparing the distribution of the high velocity $^{12}$CO (2-1) emission with the distribution of various outflow and shock tracers (CH$_{3}$OH, HCN, SiO), we discuss the interactions between the outflows and the dense molecular envelope.
Details about the observations are introduced in Section \ref{chap_obs}.
The results are presented in Section \ref{chap_result}.
In Section \ref{chap_discuss}, we discuss the role of the molecular outflow, and compare it with the role of the feedback from the radiation pressure, stellar wind, and the pressure of the ionized gas. 
A brief conclusion is provided in Section \ref{chap_conclusion}.

\section{Observations}
\label{chap_obs}
We observed multiple molecular line transitions from 2005 till 2009.
The properties of the observed molecular transitions are summarized in Table \ref{table_molecule_list}; and the observational parameters are summarized in Table \ref{table_parameters}.
We also perform high resolution observations of 3.6 cm continuum emissions.
Details about calibrations and data reduction are introduced as follows.
\subsection{The 1.3 mm band observation}
We observed the CH$_{3}$OH J=5 transitions and the $^{12}$CO (2-1) line toward G10.6-0.4  using  the Submillimeter Array (SMA)\footnote{The Submillimeter Array is a joint project between the Smithsonian Astrophysical Observatory and the Academia Sinica Institute of Astronomy and Astrophysics, and is funded by the Smithsonian Institution and the Academia Sinica. More details can be referenced in Ho, Moran, \& Lo (2004).} in the compact configuration and the very extended configuration on 2009 June 10 and 2009 July 12, respectively; and observed  the HCN (3-2) line  with the SMA in the compact configuration on 2005 June 21 and in the extended configuration on 2005 May 9.
We also observed the $^{12}$CO (2-1) lines in the SMA subcompact configuration on 2009 February 09 .
The frequency resolutions in these observations are 390 kHz ($\sim$0.5 kms$^{-1}$).

\begin{figure}
\vspace{-3cm}
\hspace{1.5cm}
\includegraphics[scale=1]{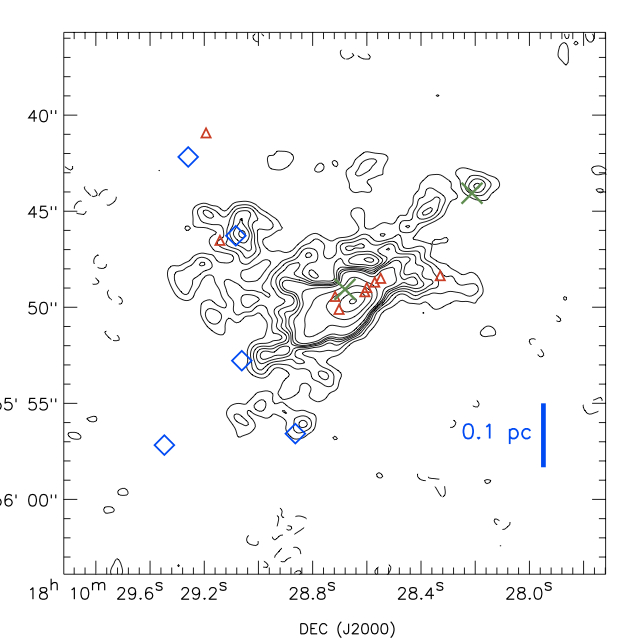}
\caption{
The 1.3 mm continuum image. 
Contours show the levels of [1, 2, 3, 4, 5, 6, 7, 8, 9, 10, 11,12,13,14,15, 16, 32, 64, 96]\% of the peak brightness temperature 49 K. 
Red triangles mark the locations of the 22 GHz water maser (Hofner \& Churchwell 1996). 
Two green crosses mark the locations of the peaks of 1.3 cm free-free continuum emission. 
Blue diamonds mark the locations of 5 broad $^{12}$CO outflow interaction signatures (see Section \ref{chap_diagnose}). 
}
\label{fig_1p3mm}
\end{figure}

The basic calibrations are carried out in \texttt{Miriad}, the self-calibration and imaging of these data are carried out in \texttt{AIPS}.
Continuum emissions are averaged from the line free channels and then subtracted from the line data.
We construct the continuum band visibility data at 1.3 mm from the line--free channels in the SMA subcompact, compact, and very--extended array data. 
We combined the data of all three array configurations, which yields a synthesized beam of  0$''$.79$\times$0$''$.58 with position angle of 60$^{o}$.
The 1.3 mm continuum image is shown in Figure \ref{fig_1p3mm}.
The discussions of this continuum image are postponed to Section \ref{chap_diagnose}.

We combine the $^{12}$CO data of all three array configurations, which yield a synthesized beam of  1$''$.2$\times$1$''$.2 with a position angle of 87$^{o}$; and we perform averaging per three velocity channels to enhance the signal--to--noise ratio.
The observed RMS noise of the $^{12}$CO (2-1) line channel maps is 0.03 Jy/beam (0.48 K) in each 1.5 kms$^{-1}$ channels, where there is no extended strong line emissions.
We combine the compact--array and the very--extended array CH$_{3}$OH data, which yield 1$''$.5$\times$1$''$.3 angular resolution and rms noise level of 0.06 Jy/beam (0.64 K) in each 0.5 km/s velocity channel.
We combine the compact--array and the extended array HCN (3-2) data, yielding 1$''$$.3\times$1$''$.1 resolution, and RMS noise of 0.24 Jy/beam (3.03 K) in each 0.5 kms$^{-1}$ channel.

\subsection{The 0.85 mm band observation}
We observed the SiO (8-7) line using the SMA in the compact configuration on 2008 October 8.
The frequency resolution in this observation is 812.5 kHz ($\sim$0.7 kms$^{-1}$).

Calibrations are carried out in \texttt{Miriad}, and imaging is carried out in AIPS.
Owing to a non-uniform uv coverage in this observation, using natural weighting, we obtain a 4$''$.9$\times$1$''$.5 synthesized beam with a position angle of 21$^{o}$.
The RMS noise in each 0.7 kms$^{-1}$ channel is 0.24 Jy/beam (0.33 K).

\subsection{3.6 cm continuum observation}
The X--band continuum emission toward G10.6-0.4 was observed in the VLA A-configuration including the VLBA Pie-Town antenna on 2005 January 2.
The sources 1331+305 and 1820-254 were observed as absolute flux and gain calibrators.
 
The basic calibrations, self-calibration, and imaging of these data are done in \texttt{AIPS} package.
The synthesized beam of the image is 0$''$.34$\times$0$''$.19 with a position angle of -7.5$^{o}$.
The observed RMS noise of the 3.6 cm continuum image is about 0.1 mJy/beam ($\sim$27 K in brightness temperature). 
We note that the 3.6 cm continuum emission is free--free emission from spectral index measurements (Keto, Zhang, \& Kurtz 2008). 

\section{Results}  
\label{chap_result}
In the following sections, we inspect the spatial distribution of the high velocity $^{12}$CO outflows, with reference to the overall distribution of the dense gas in G10.6-0.4.
We trace the dense molecular gas by the emission of the CH$_{3}$OH J=5 transitions (Liu et al. 2010 b).
The CH$_{3}$OH 5(0,5)-4(0,4) A+ transition, the CH$_{3}$OH 5(0,5)-4(0,4) E transition, and the blended CH$_{3}$OH 5(2,3)-4(2,2) E and 5(-2,4)-4(-2,3) E transitions trace an extended 0.5 pc scale massive envelope. 
The highest excitation CH$_{3}$OH 5(-3,3)-4(-3,2) E transition exclusively trace a $<$0.1 pc scale hot toroid, which is revealed as a white compact ellipse in the middle of the CH$_{3}$OH RGB image (Figure \ref{fig_co}).
As will be discussed in the position--velocity diagram (Figure \ref{fig_ch3oh_pv}), the CH$_{3}$OH emission shows an integrated rotational motion, and trace the warm and dense accretion flow around the OB cluster. 
By comparing the outflows with the CH$_{3}$OH emission, we try to emphasize the formation of a cluster of stars in the self--gravitational accretion flow.  

The high velocity outflows are interacting with the ambient dense gas.
We diagnose the interactions between the high velocity outflows and the ambient dense gas via the broad CH$_{3}$OH emission and also the distribution of the water maser sources (Hofner \& Churchwell 1996).
From the flux of the $^{12}$CO emission, we estimate the mass, energy, and momentum feedback.
We constrain the feedback rate from the outflow dynamical timescale.

In addition, we report the detection of significantly enhanced HCN (3-2) emission, which has the distribution and the velocity range that are consistent with one of the strongest blueshifted $^{12}$CO outflow components near the central hot toroid (we call this blueshifted outflow component \textit{HCN outflow} hereafter ). 
We also detect localized broad SiO (8-7) emission and geometrically elongated 3.6 cm free--free continuum emissions around  \textit{HCN outflow}.
The physical implications of these observed signatures are discussed. 

For the sake of clarity, the discussion in the following sections are mostly based on the velocity integrated maps.
The $^{12}$CO (2-1) and the HCN (3-2) channel maps are presented in Appendix \ref{chap_channel}.

\subsection{The Spatial Distribution and Morphology of the $^{12}$CO Emission}
\label{chap_name}
\begin{figure}
\vspace{10cm}
\hspace{-2.5cm}
\resizebox{!}{8cm}{
\includegraphics[scale=0.8]{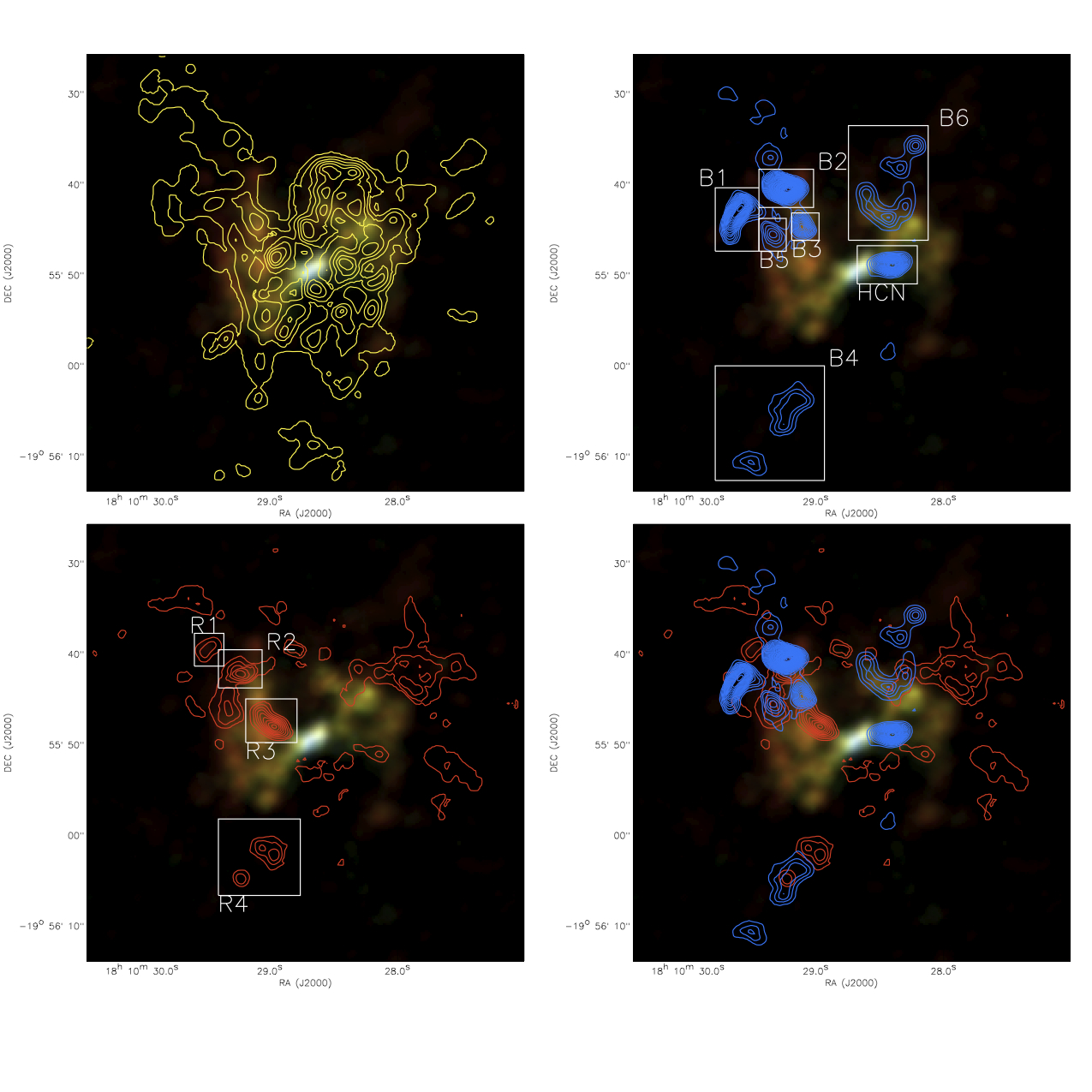}}         
\vspace{-1.5cm} 
\caption{
\textbf{Top Left: } The RGB image of the CH$_{3}$OH J=5 transitions (\textbf{R:} CH$_{3}$OH 5(0,5)-4(0,4) A+; \textbf{G:} CH$_{3}$OH 5(-2,4)-4(-2,3) E and CH$_{3}$OH 5 (2,3)-4(2,2) E; \textbf{B:} CH$_{3}$OH 5(-3,3)-4(-3,2) E), overlaid with the velocity integrated $^{12}$CO (2-1) emission in the intermediate velocity range (-15.25--8.75 kms$^{-1}$) CO. Contours start from 3 Jy/beam$\cdot$kms$^{-1}$, with 9 Jy/beam$\cdot$kms$^{-1}$ intervals. Note the CO emissions in this velocity range is suffer from missing flux and the high optical depth of the foreground gas.
\textbf{Top Right: } The RGB image of the CH$_{3}$OH J=5 transitions, overlaid with the velocity integrated $^{12}$CO (2-1) emission in the blueshifted velocity range (-124.75-- -15.25 kms$^{-1}$) CO. Contours start from 1 Jy/beam$\cdot$kms$^{-1}$, with 1 Jy/beam$\cdot$kms$^{-1}$ intervals. 
\textbf{Bottom Left: } The RGB image of the CH$_{3}$OH J=5 transitions, overlaid with the velocity integrated $^{12}$CO (2-1) emission in the redshifted velocity range (8.75-- 46.25 kms$^{-1}$) CO. Contours start from 0.5 Jy/beam$\cdot$kms$^{-1}$, with 0.5 Jy/beam$\cdot$kms$^{-1}$ intervals.
\textbf{Bottom Right: }The RGB image of the CH$_{3}$OH J=5 transitions overlaid with the both the redshifted and blueshifted $^{12}$CO (2-1)  emission. 
}
\label{fig_co}
\end{figure}

Figure \ref{fig_co} shows the velocity integrated maps of the $^{12}$CO (2-1) emission in the blueshifted, redshifted, and the intermediate velocity ranges, superposed on the dense molecular envelope as traced by the velocity integrated emission of CH$_{3}$OH J=5 transitions.
The intermediate velocity range is defined as -15.25--8.75 kms$^{-1}$.
In this range the rms noise in channel maps is significantly worse owing to the difficulties in imaging the strong and extended emission.
The blueshifted  (-124.75-- -15.25 kms$^{-1}$) and the redshifted  (8.75-- 46.25 kms$^{-1}$)  velocity ranges are defined relative to the intermediate velocity range.
The system velocity of G10.6-0.4 is -3 kms$^{-1}$, which is about the midpoint of the intermediate velocity range.
We integrate the emission in these three limited velocity ranges such that the significant features are not diluted in the velocity space.
Some highly redshifted $^{12}$CO emissions marginally detected in the velocity channels of 48.5 kms$^{-1}$-- 57.5 kms$^{-1}$ (Figure \ref{fig_cochan1}) will be discussed separately.
Outside of these selected velocity ranges, we do not see significant detection of $^{12}$CO emission.
According to the galactic rotation curve (Caswell et al. 1975; Corbel \& Eikenberry 2004), the $^{12}$CO emission in the velocity range of $v_{lsr}$= -3 -- 45 kms$^{-1}$ is potentially confused with the foreground molecular gas emission/absorption, which may explain why the velocity integrated map of the redshifted $^{12}$CO shows more diffuse emission than that of the blueshifted $^{12}$CO (see also the discussions in Appendix \ref{chap_channel}).

We do not detect the collimated high velocity bipolar outflow aligned with the rotational axis of the central hot toroid.
In the blueshifted $^{12}$CO velocity integrated map, which is not biased by the foreground emission/absorption, we see individual compact components.
These compact components have the angular size scale of a few arcseconds.
From the channel maps (Figures \ref{fig_cochan1}, \ref{fig_cochan2}, \ref{fig_cochan3}, \ref{fig_cochan4}), one can see that these compact blueshifted outflow systems have very different coverages in the velocity space.
This morphology of the high velocity $^{12}$CO emissions qualitatively look very different from the well--defined massive bipolar molecular outflow (see the case of G240.31+0.07 for an example; Qiu et al. 2009), and are difficult to be interpreted by a single wide angle wind. 
These result can either be explained by multiple powering sources, or can be explained by powerful non-uniform or episodic bursts.
These different explanations are not mutually exclusive.
Future observations with higher sensitivity and higher resolution to detect the protostellar cores and to resolve the morphology/alignment of individual outflows/jets will help to clarify the origin of the high velocity $^{12}$CO emissions.
We also note that G10.6-0.4 is a quite evolved massive cluster forming region with extremely bright H\textsc{ii} regions.
Comparing the physical size scale of the entire dense envelope ($\sim$0.5 pc) with that of Orion 
\footnote{The field--of--view of our $^{12}$CO observations in G10.6-0.4 is about 1$'$. 
The physical scales of the observed area and the resolution of this G10.6-0.4 data are comparable with those of the $^{12}$CO survey in OMC--2 and OMC--3, reported by Williams, Plambeck, \& Heyer (2003). See Takahashi et al. (2008) for a more recent study of molecular outflows in OMC--2 and OMC--3.}
, it is not surprising if multiple (proto--)stellar objects have already been formed and are widely distributed in the entire massive envelope\footnote{We define the \textit{massive envelope} as the region with molecular hydrogen density greater than 10$^{5}$ cm$^{-3}$, and with temperature greater than 30 K. Such a region can be traced by the NH$_{3}$ main hyperfine inversion line, and the CH$_{3}$OH 5(0,5)-4(0,4) A+ transition (Liu et al. 2010 b). }.
In active accretion phase, those (proto--)stellar objects will also eject molecular outflows, which will contribute to the high velocity $^{12}$CO emissions.

The size scale and the spatial distribution of the significantly detected compact redshifted $^{12}$CO emission are qualitatively consistent with the results of the blueshifted $^{12}$CO emission.
However, from the channel maps (Figures \ref{fig_cochan1}, \ref{fig_cochan2}, \ref{fig_cochan3}, \ref{fig_cochan4}), we see that the velocity coverage of the redshifted $^{12}$CO emission is about a factor of 2 smaller than the blueshifted emission.
This can be either explained by the intrinsic asymmetry in some specific powering sources, or the asymmetry of the entrained ambient molecular gas.
For the sake of the convenience of the following discussions, we label several compact blueshifted and redshifted components on the velocity integrated maps, which are visually identified from the channel maps.
When we mention the labels like R$_{i}$, B$_{i}$ outflow, or \textit{HCN outflow},  we specifically refer to the $^{12}$CO emission in the corresponding velocity ranges.

By comparing the spatial distribution of the redshifted and blueshifted $^{12}$CO, we see that the compact components from these distinct velocity ranges are highly correlated.  
The R$_{4}$ outflow and the B$_{4}$ outflow are spatially close to each other, but, are isolated from the massive envelope as traced by the CH$_{3}$OH emission.
The averaged line--of--sight velocities\footnote{The averaged line--of--sight velocities is derived by $I$/$M$, where $M$ and $I$ are the mass and the line--of--sight momentum integrated in the blueshifted or the redshifted velocity range (see Table \ref{table_outflow}.).} of R$_{4}$ outflow and the B$_{4}$ outflow have comparable absolute values, which are 14.6 kms$^{-1}$ and 18.1 kms$^{-1}$, respectively.
The B$_{4}$ outflow is elongated in the southeast--northwest direction, which is consistent with the alignment of B$_{4}$ and R$_{4}$ outflows.
These results suggest that R$_{4}$ and B$_{4}$ outflows resemble a bipolar molecular jet, and provide indirect indications that protostars form not only in the densest center of the massive envelope, but also form in other local overdensities. 
The B$_{6}$ outflow consist a \textbf{V} shaped feature, and two other compact components north of the \textbf{V} feature. 
The \textbf{V} shaped feature can also be visually identified from the $^{12}$CO channel maps, in the velocity range of -19.5-- -14.5 kms$^{-1}$.
One appealing explanation of the morphology of the $^{12}$CO emission in B$_{6}$ outflow is one collimated molecular outflow with wide angle wind.  
We  note that spatially the B$_{6}$ outflow is associated with an ultracompact HII region (see also Figures \ref{fig_ch3oh_outflow0}, \ref{fig_ch3oh_outflow2}), of which we detected the arc shaped expansional signature with -5 kms$^{-1}$$<$v$_{lsr}$$<$5 kms$^{-1}$ in the CS (1-0) position-velocity diagram (Liu et al. 2010 b).
The molecular wind, the ionized gas pressure, and the ionizing photons together may have created the \textbf{V} shaped cavity.

Northeast of the hot toroid traced by the CH$_{3}$OH 5(-3,3)-4(-3,2) E transition (E$_{u}$ = 96.93 K),  we see abundant outflows or activities related to outflows.
With the current resolution, the detailed morphology of each outflow system is poorly resolved, and from $^{12}$CO emission only it is difficult to distinguish if there is individual bipolar outflow in this region.
Some hints are provided by the water maser detections.
Water masers form in high density (n$_{H_{2}} = $10$^{8}$--10$^{9}$ cm$^{-3}$), and high temperature ($\sim$2000 K) environment, and are normally found very close to protostars (Garay \& Lizano 1999).
In Figures \ref{fig_ch3oh_outflow0} and \ref{fig_ch3oh_outflow2}, we plot the positions of water masers from Hofner \& Churchwell (1996).
From these figures we see that one maser source (RA:18$^{h}$10$^{m}$29.192$^{s}$, Decl: -19$^{o}$55$'$40.93$''$) seems to be associated with B$_{2}$ or R$_{2}$ outflow; and also one maser source (RA:18$^{h}$10$^{m}$29.144$^{s}$, Decl: -19$^{o}$55$'$46.52$''$) seems to be associated with B$_{3}$ or R$_{3}$ outflow.
The spectrum of each of these maser sources shows a cluster of components around the cloud systemic velocity, and additionally shows some higher velocity components.
These water maser results provide strong indications that the outflows detected northeast of the hot toroid are driven by local intermediate mass protostars, or even B--type massive (proto--)stars which are in active accretion phase. 
Other water maser sources are mostly associated with the hot toroid and the central UC H\textsc{ii} region.
One water maser source at RA=18$^{h}$10$^{m}$28.328$^{s}$ and Decl= -19$^{o}$55$'$48.36$''$ may be associated with the \textit{HCN outflow}.

In the velocity range of 47.75-- 58.25 kms$^{-1}$, we marginally detect compact emissions around the locations of B$_{1}$ and B$_{2}$ outflows (Figure \ref{fig_cochan1}).
If the distance of G10.6-0.4 is 6 kpc, this velocity range is not likely to be affected by the emission/absorption of the foreground gas, and the highly redshifted $^{12}$CO emissions are likely to be the counter parts of blueshifted B$_{1}$ and B$_{2}$ outflows.

In the intermediate velocity range, the $^{12}$CO emissions show tremendous structures.
We attempt to explain the results.
However, it has to be borne in mind that the $^{12}$CO emissions in this velocity range are affected by the high optical depth of the foreground gas.
In addition, missing the short spacing in the interferometric observations makes large scale structure invisible, and also makes the imaging extremely difficult.
Therefore,  the very detailed structures in individual channel maps need to be confirmed by the cross comparison with other observations. 
Discussions will be provided to argue that the missing  short spacing problem and the foreground absorption have minor effects on our following analysis.

The velocity integrated map, the mean velocity and the velocity dispersion maps of the $^{12}$CO emission in the intermediate velocity range are provided in Figure \ref{fig_comnt012}.
Based on these figures, we suggest a bimodal picture for the geometry and the dynamics.
The majority of the dense gas agrees with the 0.5 pc scale envelope detected in CH$_{3}$OH, and has the velocity gradient from southeast to northwest. 
In the envelope, a significant \textbf{V} shaped feature of outflow cavity can be seen north of the hot toroid.
Northeast of the \textbf{V} shaped feature, we detect 5$''$--10$''$ (0.15--0.3 pc) scale filamentary structures (filament $NE$, hereafter).
From the mean velocity map and the $^{12}$CO channel maps (Figure \ref{fig_cochan2}), we see that filament $NE$ has brightness temperature of a few tens of Kelvin, and covers the velocity range of -16-- -10 kms$^{-1}$, which is broad and is significantly bluer than the systemic velocity of the dense molecular envelope ( -3 kms$^{-1}$).
A population of the redshifted gas (G10--$SW$, hereafter) is further seen southwest of the hot toroid.
The morphology of G10--$SW$ is potentially filamentary and has the northeast--southwest alignment.
However, the emission from G10--$SW$ is projected against the emission from the cavity features southern to the hot toroid (Liu et al. 2010b). 
Some hints for the filamentary morphology of G10--$SW$ can be seen in the velocity dispersion map, which shows broad line emission regions with a narrowly elongated shape.
The velocities of the filament $NE$ and the G10--$SW$ together show a gradient from the northeast to the southwest. 
These filamentary structures may be originated from ambient gas influenced by massive bipolar molecular outflow or the expansional motion of the ionized gas.
The energetics of this dynamical feature can be approximately seen from the previous single dish measurement.
By observing the $^{13}$CO (2-1) line using IRAM 30m telescope, L{\'o}pez-Sepulcre et al. (2009) estimated the mass, momentum and energy of the molecular line wind in the velocity range of -11.6-- -8.6 kms$^{-1}$ and 3.9--7.9 kms$^{-1}$ to be 200 M$_{\odot}$, 1510 M$_{\odot}$kms$^{-1}$, and 12$\cdot$10$^{46}$ erg, respectively.
The filament $NE$ is also marginally detected in our CS (1-0) observations (Liu et al. 2010 b); however, both the filament $NE$ and the G10--$SW$ are not seen in the CH$_{3}$OH images (Figure \ref{fig_co}), which provides the limits on the density and the gas excitation temperature. 
The detailed physical conditions can be constrained by future sensitive observations in CO J=2-1 and J=3-2 isotopologues transitions. 
More discussions about the velocity dispersion of the intermediate velocity $^{12}$CO emission are postponed to Section \ref{chap_diagnose}.

We note that filamentary structures with the same spatial scale are also detected around the nearby massive cluster forming region, the Orion KL region (Wiseman \& Ho 1996; Wiseman \& Ho 1998).
From the results of the NH$_{3}$ observations, Wiseman \& Ho (1998) suggested a bimodal pattern of velocity.
In addition, they found the filaments appear to be fragmenting into chains of core, which are potentially the future or the current sites for low--mass star formation.

\subsection{The Mass, Energy and Momentum of the High velocity CO Outflows}
\label{chap_estimation}
For the blueshifted and the redshifted outflows, we assume optically thin emission with LTE conditions.
We estimate the mass, momentum and energy traced by the $^{12}$CO emission by adopting the gas excitation temperature of T = 100 K and the interstellar abundance ratio [$^{12}$CO]/[H2] = 10$^{-4}$.
Our estimate does not explicitly consider the inclination angle, which statistically biases the total outflow momentum lower by a factor of order 1.
The optically thin assumption on average may also bias the estimate lower by a factor of order 1.
The angular size scale of the synthesized beam of our $^{12}$CO images is 1.2$''$.
The emissions from isolated structures with angular size scale much smaller than that of the synthesized beam are potentially beam diluted and therefore cannot be detected.
In nearby clouds, the molecular outflows/jets typically have the physical size scale of a few times of 0.1 pc, which corresponds to a few times of 3$''$ at the distance of G10.6-0.4.
This typical angular size scale is greater than the 1.2$''$ synthesized beam of our $^{12}$CO images.
Therefore we do not consider the effect of beam dilution as a dominant bias in our estimates. 
The diffuse emission may not be robustly imaged in the high resolution maps.
To check this effect, we image the $^{12}$CO (2-1) emission with the subcompact--array data alone, yielding  4$''$.7$\times$2$''$.4 synthesized beam and rms noise of 0.17 Jy/beam (0.44 K) in each velocity channel of 0.5 kms$^{-1}$ width. 
By averaging the subcompact-array channel maps in the blueshifted and the redshifted velocity ranges, we do not find significantly more diffuse high velocity emission, thus suggesting that our estimates based on the 1$''$.2$\times$1$''$.2 high resolution images (Section \ref{chap_obs}) are representative of the total mass, momentum, and energy feedback.

The $^{12}$CO outflows in the intermediate velocity range may be some energetic systems with small inclination angle; and observationally, systems in this velocity range are severely confused with the global dynamics of the dense gas (Liu et al. 2010a, b).
Therefore the direct analysis of the $^{12}$CO outflow momentum and energy in the intermediate velocity range is difficult.
In addition, the $^{12}$CO emission is optically thick in this velocity range.
Without the complementary information of the $^{13}$CO or C$^{18}$O observations, the optically depth is uncertain. 
Statistically, we expect the outflow feedback from those systems with low inclination angle quantitatively has the same order of magnitudes as the feedback from outflows which have high inclination angles and are detected in the blueshifted and the redshifted velocity ranges. 
Or, there can be some less energetic outflow systems, as those detected in OMC--2 and OMC--3 (see Williams, Plambeck, \& Heyer 2003, and therein).
The contributions of such systems are not significant as compared with the energetic high velocity outflows. 

We estimate the deconvolved size scale of each outflow system by performing 2D Gaussian fitting of the velocity integrated maps.
The outflow dynamical timescale is estimated by dividing the major axis of the fitted Gaussian with the averaged velocity.
The dominant uncertainties in the estimation of the dynamical timescale are caused by the inclination, which is not measured in these observations. 
Also, owing to the limited angular resolution, we are only able to measure the dynamical timescale greater than 1700 years\footnote{This value provides the rough sense of the limit of our data. Physically the outflow velocity is uncertain, which leads to the uncertainties in the estimation of the outflow dynamical timescale.}, assuming a typical outflow velocity of $\sim$20 kms$^{-1}$.

The dynamical timescale, mass, momentum, and energy of each significant and compact outflow system in G10.6-0.4 are summarized in Table \ref{table_outflow}.
The individual outflow systems have a typical mass of $\sim$1 M$_{\odot}$, and momentum of a few tens of M$_{\odot}$kms$^{-1}$.
Since we only estimate the physical properties of the outflow systems in the blueshifted and the redshifted velocity ranges while the real outflow systems continue into the intermediate velocity range, these quantities can be underestimated.
To provide the sense of how much bias is made by considering only the limited velocity ranges, we estimate the physical properties of the \textit{HCN outflow} both in the blueshifted velocity range and in the entire velocity range with $^{12}$CO (2-1) detection.
By including the intermediate velocity range, the dense gas in the massive envelope significantly contributes to the $^{12}$CO flux, and the estimated mass should be treated as the upper limit of the molecular mass in the outflow system.
From Table \ref{table_outflow}, we see that the results with the inclusion of the intermediate velocity range has a factor of 10 higher mass, a factor of 4 higher momentum, and a factor of 2.7 higher energy. 
These results can be understood since the kinetic energy is scaled with the square of velocity, and therefore the low velocity outflow gas in the intermediate velocity range does not contribute to a lot of momentum and energy, as long as its mass is not way much higher than the high velocity components.
Thus we conclude that our estimations of the total energy and momentum feedback based on the high velocity components only is accurate up to an order of magnitude.
The estimated physical parameters of \textit{HCN outflow} (Table \ref{table_outflow}) can be uncertain owing to the strong photo--dissociation. 
However, it is unlikely that the total molecular mass integrated in the line--of--sight of \textit{HCN outflow} can be comparably high as the mass enclosed by the nearby hot toroid ($\sim$200 M$_{\odot}$).
Based on this argument, we suggest a lower limit of the [$^{12}$CO]/[H$_{2}$] abundance ratio in \textit{HCN outflow} to be $\sim$10$^{-5}$.
We expect the [$^{12}$CO]/[H$_{2}$] abundance ratio in other outflow systems to be less biased from the typical value of 10$^{-4}$.

We estimate the outflow mass, momentum, and energy in the entire blueshifted and the redshifted velocity range by integrating the flux in the whole field of view.
For the blueshifted outflows, the total mass is about 6.2 M$_{\odot}$; the total line--of--sight momentum is 163 M$_{\odot}$kms$^{-1}$; the total energy is 6.6$\cdot$10$^{46}$ erg.
For the redshifted (also including the velocity range of 46.25--58.25 kms$^{-1}$) outflows, the total mass is about 4.4 M$_{\odot}$; the total line--of--sight momentum is 109 M$_{\odot}$kms$^{-1}$; the total energy is 3.2$\cdot$10$^{46}$ erg.
The estimated mass includes all entrained and shocked gas in the outflows which can be detected in $^{12}$CO emission.
The dynamical timescale of these outflow systems has the order of 10$^{3}$--10$^{4}$ years.
In the blueshifted velocity range, the visually identified compact systems (B$_{1\cdots6}$, $HCN$--B outflows) contribute to 89\% of the total outflow mass; 84\% of the outflow momentum; and 72\% of the outflow energy.
However, in the redshifted velocity range, the visually identified compact systems (R$_{1\cdots4}$ outflows) only contribute to 31\% of the total outflow mass; 20\% of the outflow momentum; and 16\% of the outflow energy.
The small fractional contribution of the compact systems (R$_{1\cdots4}$ outflows) in the redshifted velocity range can be explained by the significant contribution of the diffuse emission (Figure \ref{fig_cochan1}) in the velocity range of 8.75 -- 46.25 kms$^{-1}$.
Those diffuse emission can be contributed by both the foreground molecular emission and the diffuse molecular outflows, which are not distinguishable with the data we currently have.
Additionally, the flux of the molecular outflows in the redshifted velocity range is also attenuated by the foreground absorption.
Since the order of magnitudes of the total mass, momentum, and energy in the redshifted velocity range are still comparable with those in the blueshifted velocity range, we suggest that the contribution of the foreground emission are roughly compensated by the attenuation of the foreground absorption.
Therefore, we accept the estimated total outflow mass, momentum, and energy in the redshifted velocity range without modeling and correcting for the foreground effect, and we argue that the relevant discussions are not seriously biased.

The total molecular mass in the region with CH$_{3}$OH detection (Figure \ref{fig_co}) is about 1000--2000 M$_{\odot}$ (Liu et al. 2010 a).
Assuming the majority of mass is moving with 2--4 kms$^{-1}$ absolute velocity (Liu et al. 2010), the momentum and kinetic energy in this system has the order of magnitudes of 2000--8000 M$_{\odot}$$\cdot$kms$^{-1}$ and 4--32$\cdot$10$^{46}$ erg.
The mass of the outflows is about a few percent of the total molecular mass in this region.
However, energy feedback from the outflows is comparable to the kinetic energy of the filament NE and G10--SW (see Section \ref{chap_name}), and is a significant fraction of the total kinetic energy of the massive envelope. 
Note the system is marginally centrifugally supported (Liu et al. 2010), and the gravitational potential energy has the same order of magnitude as the rotational kinetic energy.
If the initial turbulence is induced by the global gravitational contraction, its kinetic energy should have the same order of magnitude as the gravitational potential energy, or less than that.

Our estimations of the outflow feedback parameters are based on the $^{12}$CO (2-1) flux.
In the Orion 1S region ($\sim$450 pc from the sun), the high resolution observation of the SiO (5-4) line unveils a cluster of molecular outflows, which are not detected in the $^{12}$CO emission (Zapata et al. 2006).
Such kind of systems may be rare, and from that observation, have the dynamical ages of few hundreds of years, and have physical size scales of 460--2700 AU. 
Although our $^{12}$CO data achieve a factor of 2 lower rms noise level than the $^{12}$CO data reported in Zapata et al. (2006), the greater distance and hence more severe beam dilution makes those system non--detectable in G10.6-0.4 in $^{12}$CO.
By assuming the LTE conditions, optically thin emission, a rotational temperature of 80 K, and an abundance ratio of [SiO]/[H$_{2}$] = 10$^{-7}$, Zapata et al. (2006) estimated the total mass, momentum, and energy in those SiO outflows to be 1.155 M$_{\odot}$, 57.85 M$_{\odot}$kms$^{-1}$, and 3.03$\cdot$10$^{46}$ ergs.
However, as suggested by the authors, at least a factor of 50 enhancement of the SiO abundance with respect to the adopted value 10$^{-7}$ is required to explain the non--detection of $^{12}$CO, implying that the real outflow mass, momentum, and energy can be a factor of 50 lower than the estimated values. 
If such kind of SiO outflow systems are rare either because of the short life time in the specific phase of weak $^{12}$CO emission, or because of their formation requires unusual physical conditions, their feedback may not be significant as compared with the total outflow feedback estimated from the $^{12}$CO emission.
With small size scale, the induced protostellar turbulence also dissipates faster.

\subsection{The Diagnostics of Interaction Signatures}
\label{chap_diagnose}
While the interaction and shock signatures typically have the velocity of a few kms$^{-1}$ from the systematic velocity, in massive cluster forming regions, the CO isotopologues emission is extremely complicated in such velocity range, and is hard to be robustly imaged in interferometric observations. 
The diagnostics of the interaction signatures are therefore relying on the observations of various tracers. 
In the following sections, we introduce the observed features in CH$_{3}$OH, HCN, and SiO, as the diagnostics in G10.6-0.4.
The interaction signatures identified from the CH$_{3}$OH emissions are cross compared with the velocity dispersion map of the $^{12}$CO (2-1) emission in the intermediate velocity range, to demonstrate the robustness of the diagnostics. 
In addition, we compare the identified interaction signatures with the 1.3 mm continuum image, which is regraded as a reliable tracer of dense molecular cores.

\subsubsection{The Broad CH$_{3}$OH Emission}
\label{chap_broadch3oh}
\paragraph{G10.6-0.4}
By comparing the velocity integrated $^{12}$CO emission with the velocity integrated CH$_{3}$OH emission, we see the association of the molecular outflows with the dense envelope.
With the terminal velocity of a few tens of kilometer per second, these molecular outflows can strongly shock the ambient dense gas in the envelope.
A variety of molecules are then released from the dust grain, and their abundances in gas phase can be dramatically enhanced.
We diagnose the interactions between the molecular outflows and the ambient gas from the broad CH$_{3}$OH 5(0,5)-4(0,4) A+ emissions.
The CH$_{3}$OH molecule is considered to be one of the early--type molecules, and is abundant in the dense protostellar core.
In addition, its abundance can be further shock enhanced by one or two orders of magnitude.
The selected transition has the lowest upper--level--energy (34.65 K) among all observed CH$_{3}$OH lines, so that the excitation is least sensitive to the stellar heating.

\begin{figure}
\hspace{-0.5cm}
\resizebox{!}{15cm}{
\includegraphics[scale=0.45]{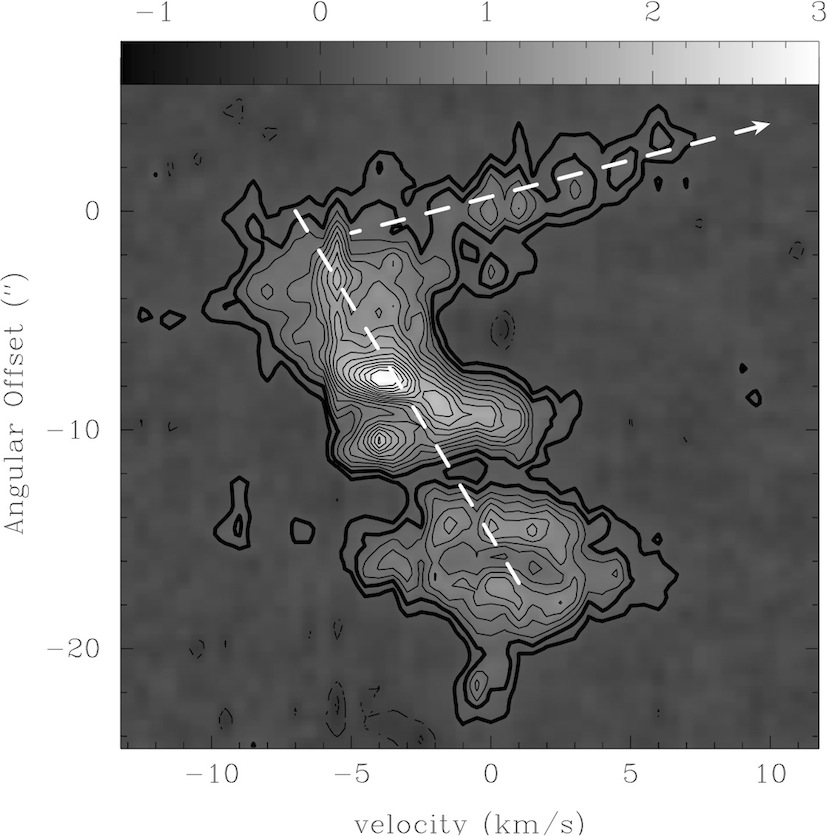}}
\caption{The pv diagram of the CH$_{3}$OH 5(0,5)-4(0,4) A+ transition. The pv cut center at ra:18$^{h}$10$^{m}$29.163$^{s}$, decl: -19$^{o}$55$'$54.65$''$, with the position angle of 140$^{o}$. Contours start from 0.18 Jy/beam (3$\sigma$) with 0.18 Jy/beam intervals; the first and the second contours are emphasized by thicker line. The dashed line indicates the general rotation, and the dashed arrow indicates a significantly broad linewidth component.}
\label{fig_ch3oh_pv}
\end{figure}

Figure \ref{fig_ch3oh_pv} shows a sample position-velocity (pv) diagram of the CH$_{3}$OH 5(0,5)-4(0,4) A+ transition.
From this figure we see two dominant signatures with distinguishable characteristics: the dense rotating envelope and the broad interaction signature. 
The dense envelope covers the angular scale of $\sim$20$''$, and is characterized by high brightness and steep increase of brightness.
The interaction signature, as indicated by the red arrow, is faint and is spatially localized, but covers a much larger velocity range.
This broad and faint emission can be understood as the outflow wing component in the spectrum. 
More pv diagrams of the interaction signatures are discussed in Appendix \ref{chap_outflowpv}. 
Qualitatively, one can observe that, in the same range of angular offset, the envelope and the interaction signature roughly occupy the same area (in unit of arcsecond$\times$kms$^{-1}$) in the pv diagram, which is naturally explained by their difference in linewidth.
However, the envelope is sketched by significantly more contours, which means that the area between two contour levels is much smaller.

The difference between the dense envelope and the interaction signature leads to a large contrast if we compare the velocity integrated flux between two significant contour levels, for example, the 3rd $\sigma$ and the 6th $\sigma$ significance levels ($f_{36}$ hereafter).
Between these two contour levels, the interaction signatures contribute to significantly more $f_{36}$ than the dense envelope, although having lower total integrated flux.
This property can be utilized to indicate the spatial distribution of the interaction signatures.
Figure \ref{fig_ch3oh_outflow0} shows the $f_{36}$ map of the CH$_{3}$OH 5(0,5)-4(0,4) A+ transition together with the locations of the water maser sources and the locations of the 1.3 cm free-free emission peaks. 

From Figure \ref{fig_ch3oh_outflow0} we see that this analysis is quite successful.
The 0.1 pc scale hot toroid (Liu et al. 2010 a,b) which has the highest brightness temperature, cannot be seen in this map.
Instead, we see significant local components (indicated by boxes) closely associated with the compact $^{12}$CO emission (for example, B$_{2}$, R$_{1}$, R$_{3}$ outflows, and B$_{6}$ outflow), which are the most likely sources to induce the interaction signatures.
We note that the CH$_{3}$OH emission is only significantly detected in the velocity range of -13 kms$^{-1}$$<$v$_{lsr}$$<$10 kms$^{-1}$ while the blueshifted and the redshifted $^{12}$CO emissions are defined outside this velocity range.
The consistency in their spatial distributions also indicates that the broad CH$_{3}$OH emission is indeed related to the high velocity molecular outflows.
We also evaluate the velocity dispersion of the emission between the 3rd $\sigma$ and the 6th $\sigma$ significance levels (Figure \ref{fig_ch3oh_outflow2}).
The result consistently shows large velocity dispersion closely associated with the interaction signatures.
Both Figure \ref{fig_ch3oh_outflow0} and Figure \ref{fig_ch3oh_outflow2} show that the interaction signature around the B6 outflow has a elongated distribution parallel to the global rotational axis of the dense envelope (Liu et al. 2010 a,b), and may be associated with an UC H\textsc{ii} region, indicated by a star symbol (Liu et al. 2010 b).
Whether that signature is related to the bipolar molecular outflow or is pushed by the bipolar expansion of the ionized gas can be studied in future high resolution observations.
The interaction signature indicated by the orange box corresponds to the interaction signature shown in the sample pv diagram (Figure \ref{fig_ch3oh_pv}).
The corresponding $^{12}$CO emission may be in the intermediate velocity range, and as shown in Figure  \ref{fig_ch3oh_outflow0}, is elongated in the southeast--northwest direction. 
It can also be visually identified in the channel map (Figure \ref{fig_cochan2}) as compact components in the velocity range of 3.5--8.0 kms$^{-1}$.

The f$_{36}$ analysis, especially, the velocity dispersion map (Figure 6) unveils a population of interaction signatures, which lie close to the plane of the densest flattened molecular gas (Liu et al. 2010 a, b).
Those interaction signatures show locality and relatively low outflow velocities.
We suggest that they are less likely due to the interaction of the wind emanated from the OB cluster embedded in the hot toroid, but rather associated with local sites of star formation.
This result provides important indication that local star formation plays important roles in the feedback of kinetic energy to the dense molecular gas.

\begin{figure}
\hspace{-1cm}
\resizebox{!}{15cm}{
\includegraphics[scale=0.8]{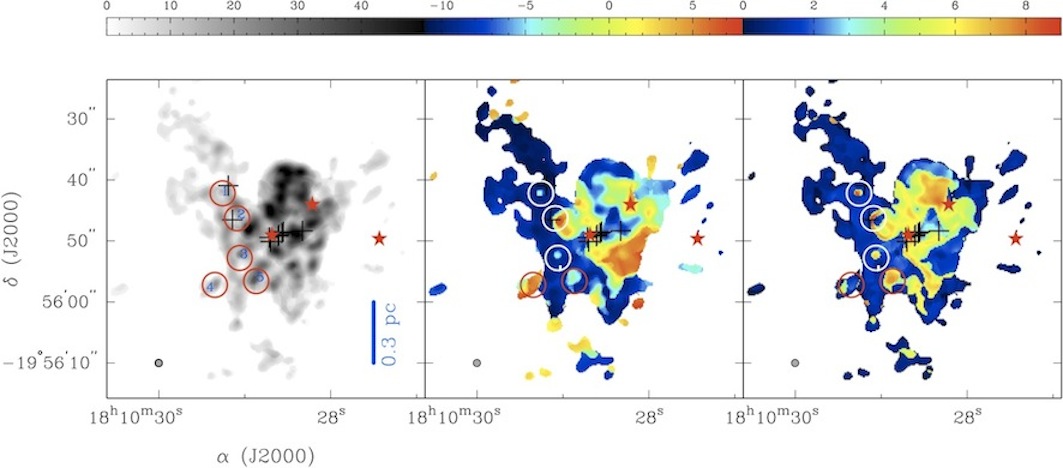}}
\caption{
\textbf{Left: } The velocity integrated map of the $^{12}$CO (2-1) line in the intermediate velocity range  (-15.25--8.75 kms$^{-1}$) . Color--bar has the unit of Jy/beam$\cdot$kms$^{-1}$.
\textbf{Middle: } The intensity--weighted mean velocity of the $^{12}$CO (2-1) emission in the intermediate velocity range  (-15.25--8.75 kms$^{-1}$). Color--bar has the unit of kms$^{-1}$. 
\textbf{Right: } The velocity dispersion map of the $^{12}$CO (2-1) emission in the intermediate velocity range  (-15.25--8.75 kms$^{-1}$). Color--bar has the unit of kms$^{-1}$.
The synthesized beam of the $^{12}$CO map is shown in the bottom left corner of each panel.
Five visually identified regions with broad line emission are marked by circles. The colors of the circles are just for better contrast in presentation, and does not have physical meaning. The coordinates of the circles can be referenced in Section \ref{chap_diagnose}.
Three 1.3 cm free-free continuum peaks (RA:18$^{h}$10$^{m}$28.683$^{s}$, Decl:-19$^{o}$55$'$49$''$.07; RA:18$^{h}$10$^{m}$28.215$^{s}$, Decl:-19$^{o}$55$'$44$''$.07 ; RA:18$^{h}$10$^{m}$27.435$^{s}$, Decl:-19$^{o}$55$'$44$''$.67) are marked by red stars. 
Cross symbols mark the water maser detections (Hofner and Churchwell 1996). 
The relative positional accuracy of the maser data is about 0.1$''$, which is much smaller than the size of the crosses.
}
\label{fig_comnt012}
\end{figure}

\begin{figure}
\resizebox{!}{15cm}{
\includegraphics[scale=0.88]{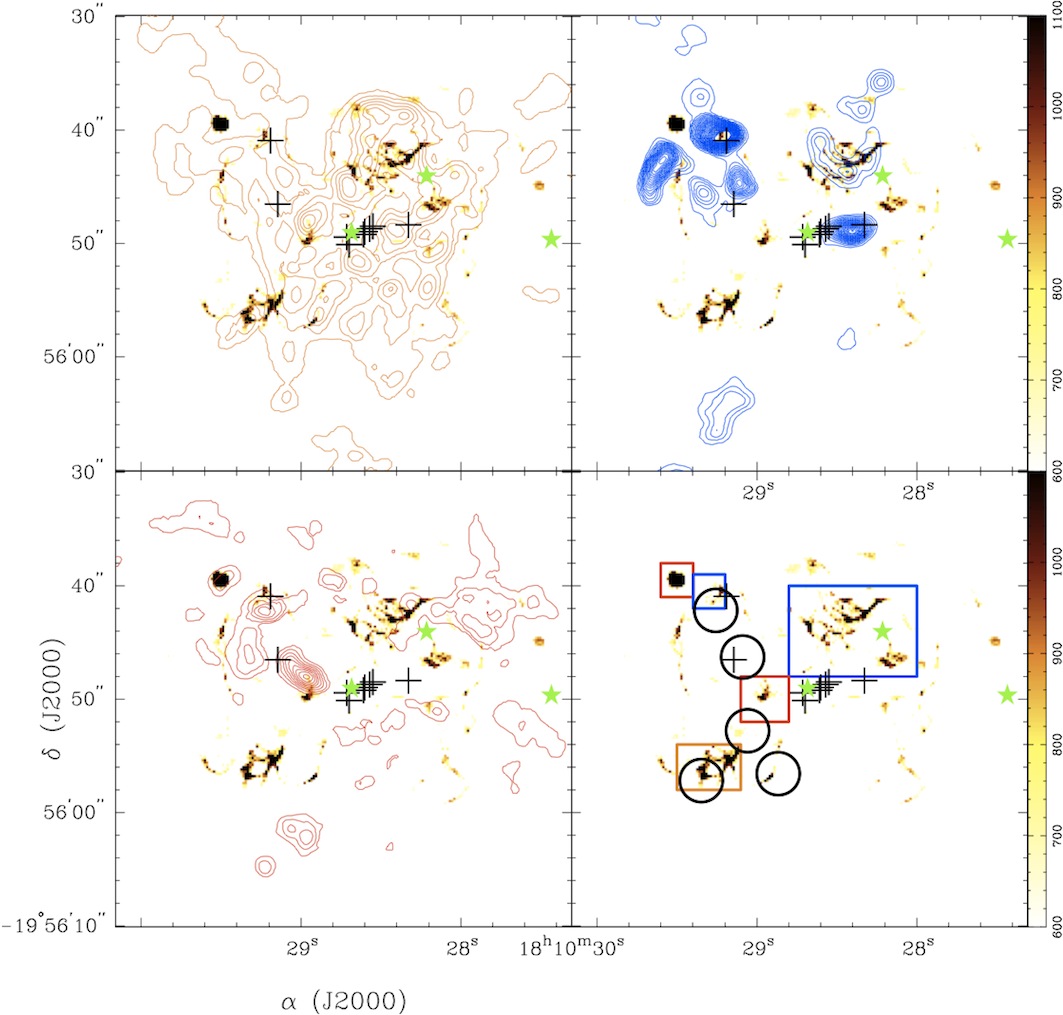}}
\caption{
\textbf{Top Left: } Intermediate velocity (-15.25--8.75 kms$^{-1}$) CO moment 0 maps. 
\textbf{Top Right: } Blueshifted high velocity (-124.75-- -15.25 kms$^{-1}$) CO (2-1) moment 0 maps.  
\textbf{Bottom Left: } Redshifted high velocity (8.75-- 46.25 kms$^{-1}$) CO (2-1) moment 0 maps. 
\textbf{Bottom Right: } The moment 0 map of low intensity emissions of CH$_{3}$OH 5(0,5)-4(0,4) A+ transition. This moment 0 map is generated by trim the data under 3$\sigma$ noise level and above 6$\sigma$ noise level. 
The bright regions in the bottom right panel are marked by rectangles with color corresponding to its CO counter part of certain contour colors. 
Three 1.3 cm free-free continuum peaks (RA:18$^{h}$10$^{m}$28.683$^{s}$, Decl:-19$^{o}$55$'$49$''$.07; RA:18$^{h}$10$^{m}$28.215$^{s}$, Decl:-19$^{o}$55$'$44$''$.07 ; RA:18$^{h}$10$^{m}$27.435$^{s}$, Decl:-19$^{o}$55$'$44$''$.67) are marked by green stars. 
Cross symbols mark the water maser detections (Hofner and Churchwell 1996). 
The relative positional accuracy of the maser data is about 0.1$''$, which is much smaller than the size of the crosses.
Circles in the left most panel are the same with those in Figure \ref{fig_comnt012}.
}
\label{fig_ch3oh_outflow0}
\end{figure}

\begin{figure}
\resizebox{!}{15cm}{
\includegraphics[scale=0.88]{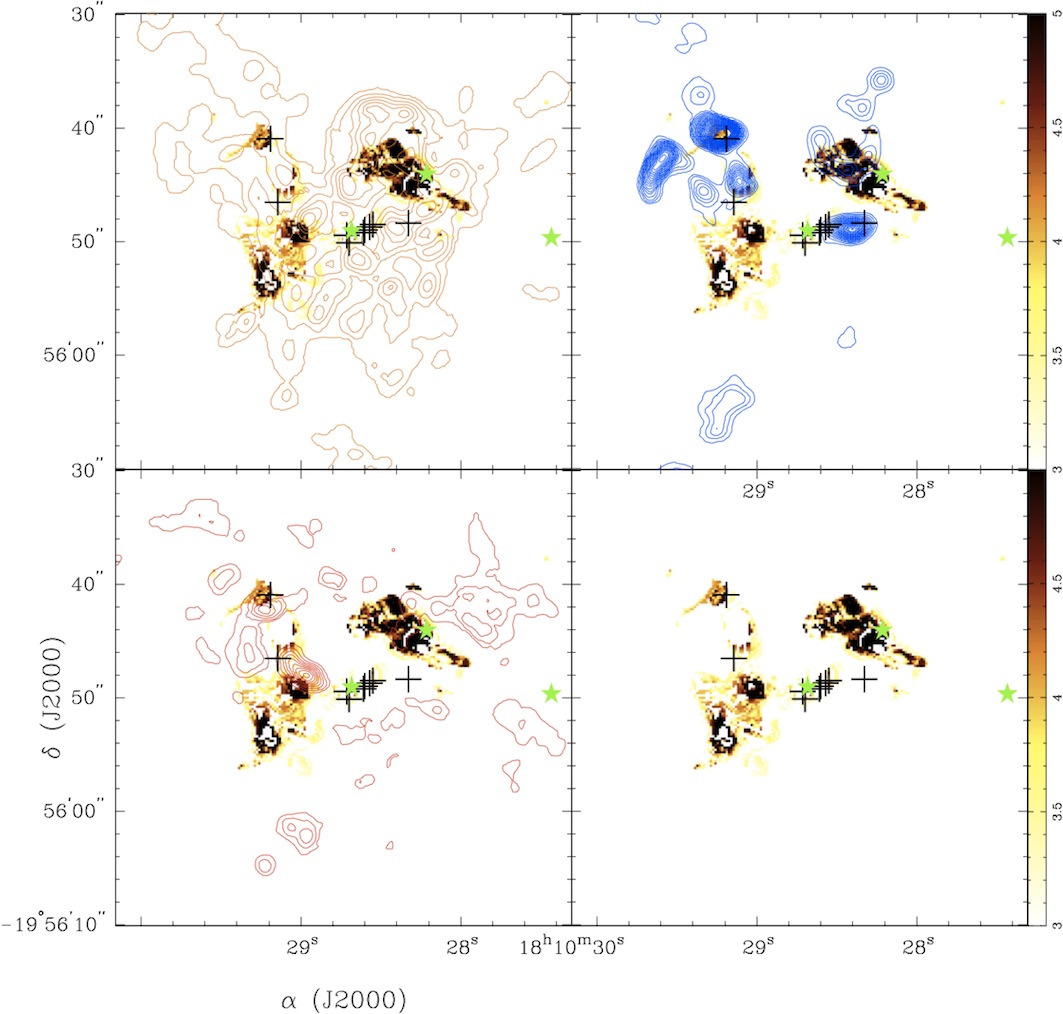}}
\caption{
\textbf{Top Left: } Intermediate velocity (-15.25--8.75 kms$^{-1}$) CO moment 0 maps.  
\textbf{Top Right: } Blueshifted high velocity (-124.75-- -15.25 kms$^{-1}$) CO (2-1) moment 0 maps. 
\textbf{Bottom Left: } Redshifted high velocity (8.75-- 46.25 kms$^{-1}$) CO (2-1) moment 0 maps. 
\textbf{Bottom Right: } The moment 2 map of low intensity emissions of CH$_{3}$OH 5(0,5)-4(0,4) A+ transition. This moment 2 map is generated by trimming the data under 3$\sigma$ noise level and above 6$\sigma$ noise level. 
The color bar has the unit of kms$^{-1}$.
Three 1.3 cm free-free continuum peaks (RA:18$^{h}$10$^{m}$28.683$^{s}$, Decl:-19$^{o}$55$'$49$''$.07; RA:18$^{h}$10$^{m}$28.215$^{s}$, Decl:-19$^{o}$55$'$44$''$.07 ; RA:18$^{h}$10$^{m}$27.435$^{s}$, Decl:-19$^{o}$55$'$44$''$.67) are marked by green stars.
Cross symbols mark the water maser detections (Hofner and Churchwell 1996). 
The relative positional accuracy of the maser data is about 0.1$''$, which is much smaller than the size of the crosses.}
\label{fig_ch3oh_outflow2}
\end{figure}

\paragraph{Cross Comparison with $^{12}$CO Emission in the Intermediate Velocity Range}
In this section, we cross compare the broad CH$_{3}$OH 5(0,5)-4(0,4) A+ emission (Figure \ref{fig_ch3oh_outflow0}, \ref{fig_ch3oh_outflow2}) with the broad $^{12}$CO emission in the intermediate velocity range (Figure \ref{fig_comnt012}). 
From the velocity dispersion map of the intermediate velocity $^{12}$CO emission (Figure \ref{fig_comnt012}, right), one can see two extended broad line regions, which are associated with two UC H\textsc{ii} regions (H\textsc{ii}--M: RA=18$^{h}$10$^{m}$28.683$^{s}$, Decl= -19$^{o}$55$'$49$''$.07; H\textsc{ii}--NW: RA=18$^{h}$10$^{m}$28.215$^{s}$, Decl=-19$^{o}$55$^{'}$44$''$.07), respectively.
The physical size scales of these two broad emission regions are both a fraction of a parsec. 

The distribution of the broad $^{12}$CO emission around H\textsc{ii}--NW agrees well with the interaction signature traced by the $f_{36}$ of the CH$_{3}$OH 5(0,5)-4(0,4) A+ transition (Figure \ref{fig_ch3oh_outflow0}). 
The morphology can be explained by the expansional motion of the molecular/ionized gas confined by the dense gas, which has a flattened distribution in the global plane of rotation (Liu et al. 2010 a b).
The  broad $^{12}$CO emission around H\textsc{ii}--M does not have CH$_{3}$OH 5(0,5)-4(0,4) A+ emission counter part, which can be explained by the lower optical depth around this specific region.

In addition to those two extended broad $^{12}$CO emission regions, we visually identify five very compact (1$''$--2$''$) broad $^{12}$CO emission signatures, for which the mean velocities also have significant contrasts with the ambient gas.
Those regions are marked by circles in Figure \ref{fig_comnt012}, and are also marked in the left most panel of Figure \ref{fig_ch3oh_outflow0}.
The central coordinates of these five circles are: (\#1) RA=18$^{h}$10$^{m}$29.258$^{s}$, Decl= -19$^{o}$55$'$42$''$.18; (\#2) RA=18$^{h}$10$^{m}$29.088$^{s}$, Decl= -19$^{o}$55$'$46$''$.28; (\#3) RA=18$^{h}$10$^{m}$29.059$^{s}$, Decl= -19$^{o}$55$'$52$''$.78; (\#4) RA=18$^{h}$10$^{m}$29.349$^{s}$, Decl= -19$^{o}$55$'$57$''$.18; (\#5) RA=18$^{h}$10$^{m}$28.868$^{s}$, Decl= -19$^{o}$55$'$56$''$.58, respectively (see also the channel maps in Figure \ref{fig_cochan2}).
Two of them (\#1, \#2) are closely associated with water maser sources.
The east most one (\#4) is associated with a $f_{36}$ peak of the CH$_{3}$OH 5(0,5)-4(0,4) A+ transition, which is marked by the yellow box in Figure \ref{fig_ch3oh_outflow0}.
A faint $f_{36}$ peak of the CH$_{3}$OH 5(0,5)-4(0,4) A+ transition is detected around \#3; and similarly with \#5.
These five signatures marked by circles look compact even from the $^{12}$CO (2-1) transition, which is very easy to be excited.
It provides the indication that they are associated with active local dynamics.
Since their mean velocities are offset from the systemic velocity of their ambient gas by a few kms$^{-1}$, we hypothesize that those signatures are the molecular gas in the local dense cores pushed by the protostellar outflows. 
Without the comparison with results from the outflow tracers, it is extremely difficult to recognize these interaction signatures from the $^{12}$CO (2-1) maps, and it is even more difficult to robustly interpret them when they are recognized.  

We note that the missing flux and the high optical depth of the foreground gas can artificially enhance the measured velocity dispersion. 
However, the broad $^{12}$CO emission regions mentioned in this section have much smaller angular size scales ($<$20$''$) than the maximum detectable angular scale of our SMA subcompact--array data ($\sim$40$''$), and should not be severely affected by missing flux.
If the missing flux artificially enhances the measured velocity dispersion at large scale, the contrast between the measured velocity dispersion of the broad $^{12}$CO emission regions and the measured velocity dispersion of the envelope is reduced. 
The contrast in velocity dispersion should be higher in reality, which means the identified broad emission signatures are robust. 
Therefore we think the missing flux does not have a significant impact on our discussions in this section.
The high optical depth of the foreground gas trims structures for all angular size scales, however, only for the emissions redder than the systemic velocity of -3 kms$^{-1}$. 
The mean velocities in those extended and compact broad $^{12}$CO emission regions are apparently redder than the mean velocity of the ambient gas.
If the high velocity dispersion is an artifact caused by the high optical depth of the foreground gas, we expect to detect a bluer mean velocity.

\paragraph{Compare With 1.3 mm Continuum Emissions}
In massive cluster forming regions, flux in 1.3 mm continuum emission is dominantly contributed by the thermal dust emission, and the free--free continuum emission from UC H\textsc{ii} regions (Keto, Zhang, \& Kurtz 2008).
To see the relation between those broad $^{12}$CO interaction signatures with the dense molecular gas, we mark the locations of those broad $^{12}$CO interaction signatures (as well as the locations of the 22 GHz water masers) on the 1.3 mm continuum image (Figure \ref{fig_1p3mm}).

From the 1.3 mm continuum image, we first see an abrupt increase of the 1.3 mm continuum flux in the middle, which suggests the distinct origins of the flux .
We identify two UC H\textsc{ii} regions from high resolution centimeter continuum images (Keto, Ho \& Haschick 1988; Guilloteau et al. 1988; Sollins et al. 2005; Liu et al. 2010abc), and mark their locations on the 1.3 mm continuum image.
Those centimeter continuum emissions indicate that in the middle of the 1.3 mm continuum map, the free--free continuum emission from the UC H\textsc{ii} region contribute significantly.
In the extended region, the 1.3 mm continuum emission is dominantly contributed by the thermal dust emission, except for a fainter UC H\textsc{ii} region in the northwest.

The thermal dust emission unveils abundant dense cores over a 0.5 pc region.
Three dusty dense cores ( RA:18$^{h}$10$^{m}$29$^{s}$.064, Decl: -19$^{o}$55$^{'}$46$^{''}$.2; RA:18$^{h}$10$^{m}$29$^{s}$.035, Decl:-19$^{o}$55$^{'}$49$^{''}$.7; RA:18$^{h}$10$^{m}$28$^{s}$.975, Decl: -19$^{o}$55$^{'}$52$^{''}$.5) are closely associated with the identified broad interaction signatures of $^{12}$CO outflows.
The core at the northeast (RA:18$^{h}$10$^{m}$29$^{s}$.064, Decl: -19$^{o}$55$^{'}$46$^{''}$.2) is further associated with a water maser source (Hofner \& Churchwell 1996).
The associations with water maser sources and outflow signatures indicates that those dense cores are actively forming stars.
This core shows a complicated geometry, which may be explained by hierarchical fragmentations. 
The dense core at RA= 18$^{h}$10$^{m}$28$^{s}$.975 and Decl=-19$^{o}$55$^{'}$52$^{''}$.5 is apparently elongated, and has internal structures, which may be explained by fragmentation, or can be observationally caused by blending.

\paragraph{Compare With Low--Mass Protostellar Outflows}
From the observations of the CH$_{3}$OH J=2 transitions with upper--level--energy E$_{u}$ of 4.64--12.2 K, Takakuwa, Ohashi \& Hirano (2003) suggests that in the Class 0 low mass protostar IRAM 04191+1522, the enhanced broad CH$_{3}$OH emission trace the interactions between the protostellar outflow and its parent molecular core.
Similar to what is observed in G10.6-0.4, the pv diagram of CH$_{3}$OH in IRAM 04191+1522 shows the distinguishable interaction signature with the envelope component.
The interaction signatures in IRAM 04191+1522  have the size scale of $\sim$0.03 pc, closely associated with the protostellar envelope.
Observations in other nearby protostellar outflows (Bachiller et al. 1995, 1998, and 2001; Garay et al. 1998) show the localized enhancement of the CH$_{3}$OH abundance in the strongly shocked regions around 0.1 pc from the ejecting sources.
This 0.03--0.1 pc size scale correspond to the angular scale of 1$''$--3$''$ at the distance of G10.6-0.4 (6 kpc), which is marginally resolved in our observations (Table \ref{table_parameters}); and this scale is small as compared with the size scale of the massive envelope traced by CH$_{3}$OH (20$''$--30$''$).

Supposedly these observed phenomena in nearby clouds and their interpretations can be extrapolated to the distant massive dense envelopes with higher temperature and high density.
Then each of the detected broad CH$_{3}$OH interaction signature may be associated with an independent ejecting protostar.
The distribution of the observed broad CH$_{3}$OH emission features in G10.6-0.4 suggests the abundant star formation activities over the entire $\sim$0.5 pc scale dense envelope.
In G10.6-0.4, we already detected three UC H\textsc{ii} regions, indicating that multiple massive stars have already been formed.
It is not surprising if there are many massive or low--mass (proto--)stars, which are still in accretion phase and are ejecting the molecular outflows.
We expect higher resolution observations in lower excitation CH$_{3}$OH to reveal more protostellar objects.

Note that the detection of the thermal radio jets is regarded as one of the most robust methods to identify protostellar objects in the accretion phase.  
However, in UC H\textsc{ii} regions, the free-free emission around the embedded OB cluster has the flux of a few Jy, which is of 3--4 orders of magnitude brighter than the typically flux of the thermal radio jets. 
The bright free-free emission associated with the OB cluster ionization can have complicated spatial distribution and geometry, which will confuse the detection of the thermal radio jets.
The shock enhanced molecular emission potentially provides good complementary information in the diagnostic of the star forming activities.
From Figure \ref{fig_ch3oh_outflow0}, we see that a population of protostellar and stellar activities (four CH$_{3}$OH outflow interaction signatures as indicated by three boxes and one white circle, and two UC H\textsc{ii} regions as indicated by stars) seem to have a coplanar distribution, suggesting a scenario of enhanced star formation in the rotationally flattened dense envelope.
This scenario is supported by recent numerical hydrodynamical simulations (Peters et al. 2010).  

Some protostellar and stellar activities seem to follow the filaments detected in the intermediate velocity $^{12}$CO emissions.
Owing to the projection effect, we cannot robustly distinguish whether those activities are just distributed in the bulk of the envelope.

\subsubsection{The \textit{HCN Outflow}}
\label{chap_hcnoutflow}
\begin{figure}
\resizebox{!}{15cm}{
\includegraphics[scale=1]{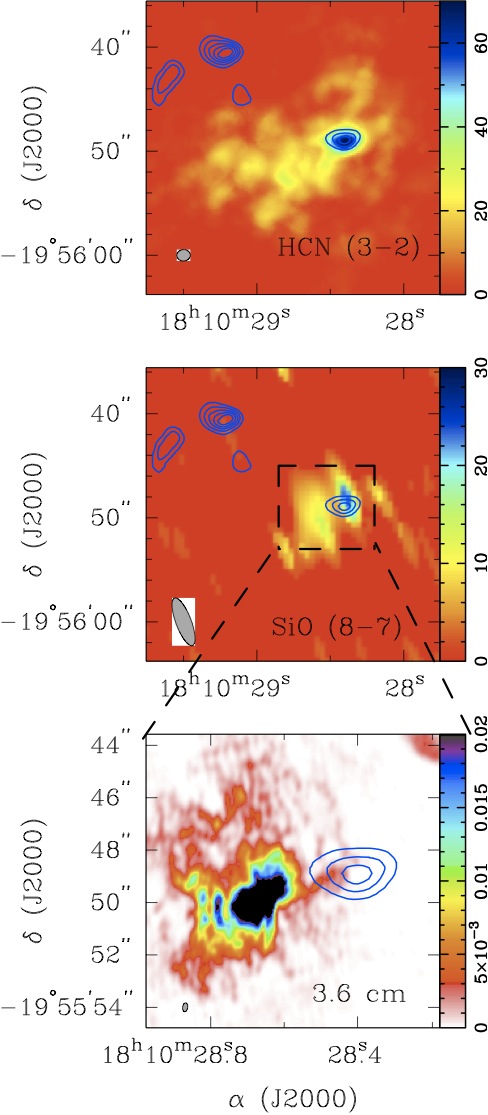}}
\caption{The HCN (3-2) moment 0 map (Top), the SiO (8-7) moment 0 map (Middle), and the 3.6 cm continuum map (Bottom), overlaid with the blueshifted high velocity (-124.75-- -15.25 kms$^{-1}$) CO (2-1) moment 0 maps (Contour). 
The units of the color bars are Jy/beam$\cdot$kms$^{-1}$. Contours start from 6 Jy/beam$\cdot$kms$^{-1}$, with 6 Jy/beam$\cdot$kms$^{-1}$ intervals. The synthesized beams of the HCN, SiO, and the 3.6 cm continuum observations are shown in the bottom left corner of these panels.}
\label{fig_hcn}
\end{figure}

The \textit{HCN outflow}  is characterized by the significantly higher HCN (3-2) flux.
Among all outflow systems detected by $^{12}$CO, the \textit{HCN outflow} is also the unique case where we detect the SiO (8-7) emission.
The deconvolved size scale of the $^{12}$CO emission is about 2$''$, and its location is close to the central 0.1 pc scale hot toroid. 
Figure \ref{fig_hcn} shows the velocity integrated maps of HCN (3-2), SiO (8-7), and the blueshifted $^{12}$CO, in the top and the middle panels, and shows the 3.6 cm continuum image in the bottom panel.
From this figure we see the locally enhanced HCN emission, for which the location and geometry agree excellently with those of its $^{12}$CO counter part.
The distribution of the SiO emission is consistent with the $^{12}$CO and HCN emission.
Subjected to the non--uniform uv coverage, the SiO (8-7) data have an elongated synthesized beam, and therefore the projected geometry of the emission and the peak location are poorly constrained. 

\begin{figure}
\resizebox{!}{15cm}{
\includegraphics[scale=0.9]{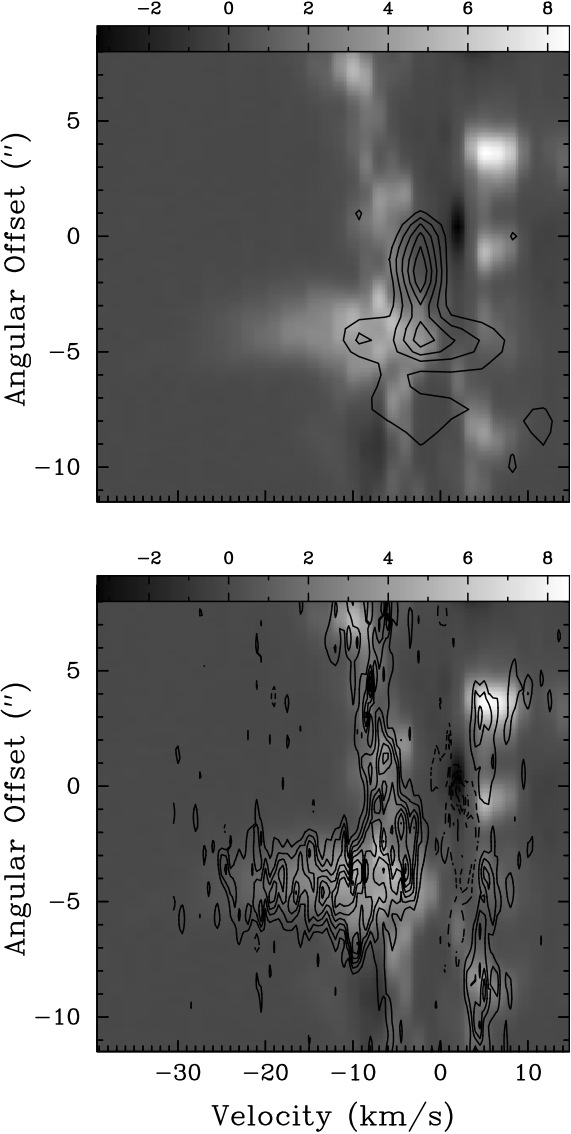}}
\caption{The pv diagrams of $^{12}$CO (2-1) (grey scale), SiO (8-7), and HCN (3-2) around the \textit{HCN outflow}.
The pv cut is centered at (RA=18$^{h}$10$^{m}$28.683$^{s}$, Decl = -19$^{o}$55$''$49.07$'$), with position angle of 90$^{o}$.
The positive angular offset is defined in the east. 
 Contours in the top panel represents the SiO (8-7) emission, start from 0.48 Jy/beam with 0.48 Jy/beam intervals.
 Contours in the bottom panel represents the HCN (3-2) emission and absorption; positive contours start from 0.72 Jy/beam with 0.72 Jy/beam intervals; negative contours start from -0.72 Jy/beam with -0.72 Jy/beam intervals. Before performing the pv cut, the SiO (8-7) image cube is \texttt{Hanning smoothed} by 3.5 kms$^{-1}$ width to enhance the signal--to--noise ratio.}
\label{fig_hcnpv}
\end{figure}

\begin{figure}
\resizebox{!}{15cm}{
\includegraphics[scale=0.8]{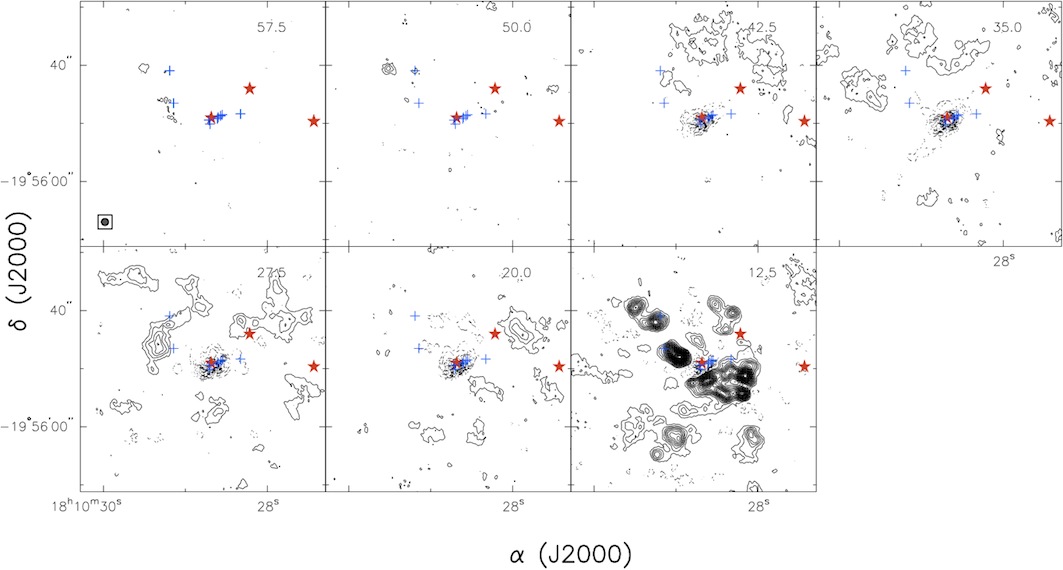}}
\caption{The channel maps of the $^{12}$CO (2-1) line.  We bin per five 1.5 kms$^{-1}$ velocity channels into one 7.5 kms$^{-1}$ velocity channel.
Solid contours start from 0.045 Jy/beam with 0.045 Jy/beam intervals; dashed contours start from -0.045 Jy/beam with -0.045 Jy/beam intervals. 
Three 1.3 cm free-free continuum peaks (RA:18$^{h}$10$^{m}$28.683$^{s}$, Decl:-19$^{o}$55$'$49$''$.07; RA:18$^{h}$10$^{m}$28.215$^{s}$, Decl:-19$^{o}$55$'$44$''$.07 ; RA:18$^{h}$10$^{m}$27.435$^{s}$, Decl:-19$^{o}$55$'$44$''$.67) are marked by red stars. Cross symbols mark the water maser detections (Hofner \& Churchwell 1996).
}
\label{fig_cochan1}
\end{figure}

\begin{figure}
\resizebox{!}{15cm}{
\includegraphics[scale=1.05]{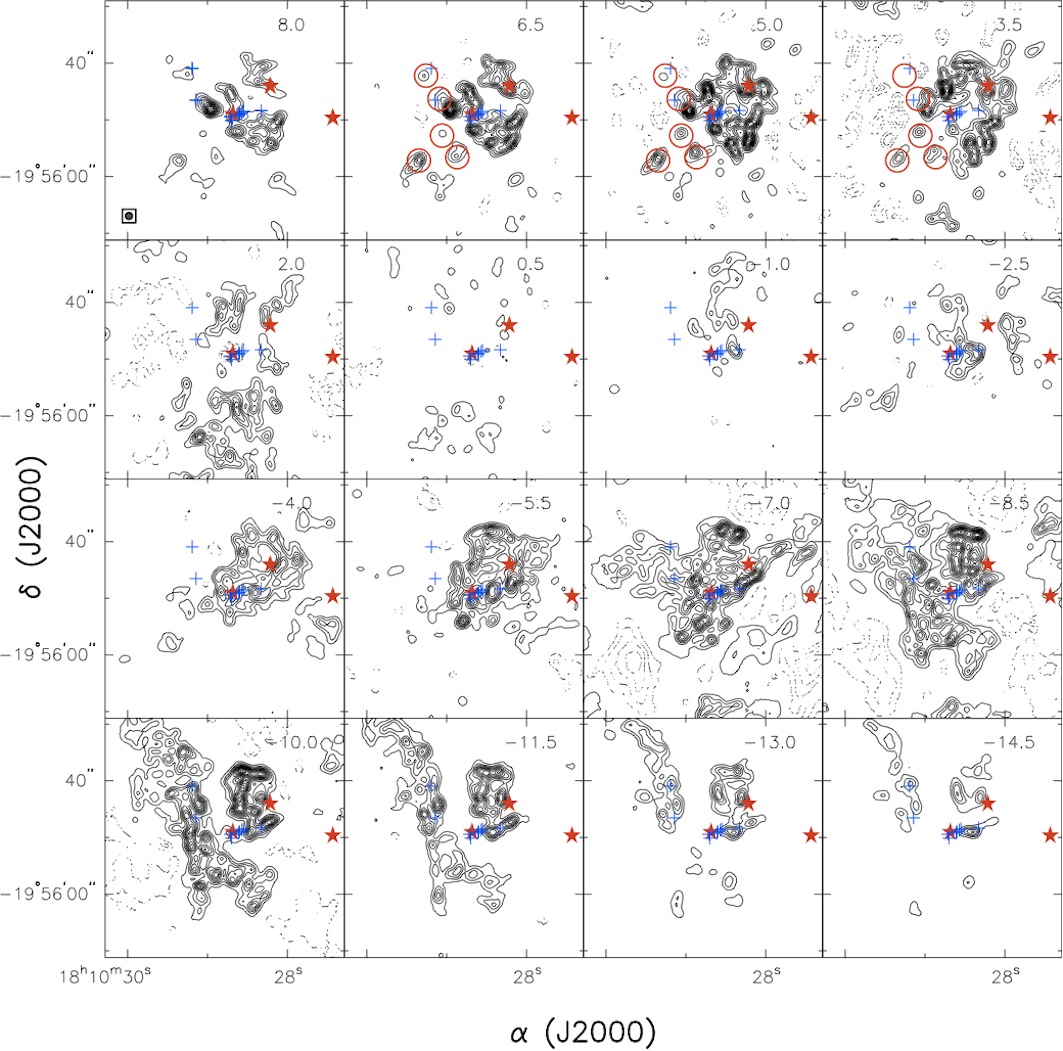}}
\caption{The channel maps of the $^{12}$CO (2-1) line.  
Solid contours start from 0.7 Jy/beam (11.2 K) with 0.7 Jy/beam intervals; dashed contours start from -0.7 Jy/beam with -0.7 Jy/beam intervals. 
Five visually identified regions with broad line emission are marked by circles. The colors of the circles are just for better contrast in presentation, and does not have physical meaning. The coordinates of the circles can be referenced in Section \ref{chap_diagnose}.
Five visually identified regions with broad line emission are marked by circles (see also Figure \ref{fig_comnt012}). 
}
\label{fig_cochan2}
\end{figure}

\begin{figure}
\resizebox{!}{15cm}{
\includegraphics[scale=0.8]{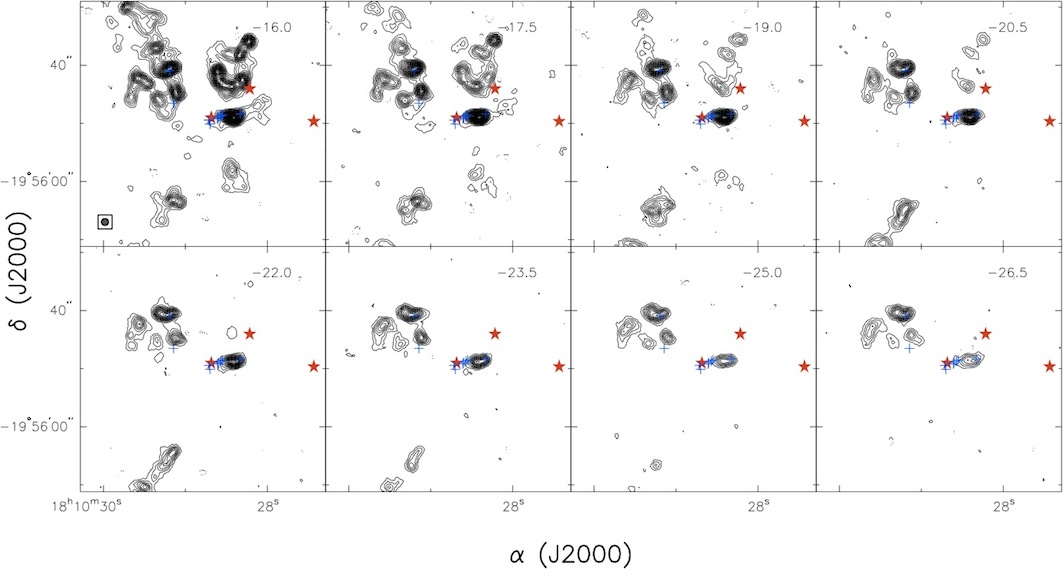}}
\caption{The channel maps of the $^{12}$CO (2-1) line.  
Solid contours start from 0.09 Jy/beam with 0.09 Jy/beam intervals; dashed contours start from -0.09 Jy/beam with -0.09 Jy/beam intervals. 
}
\label{fig_cochan3}
\end{figure}

\begin{figure}
\resizebox{!}{15cm}{
\includegraphics[scale=0.8]{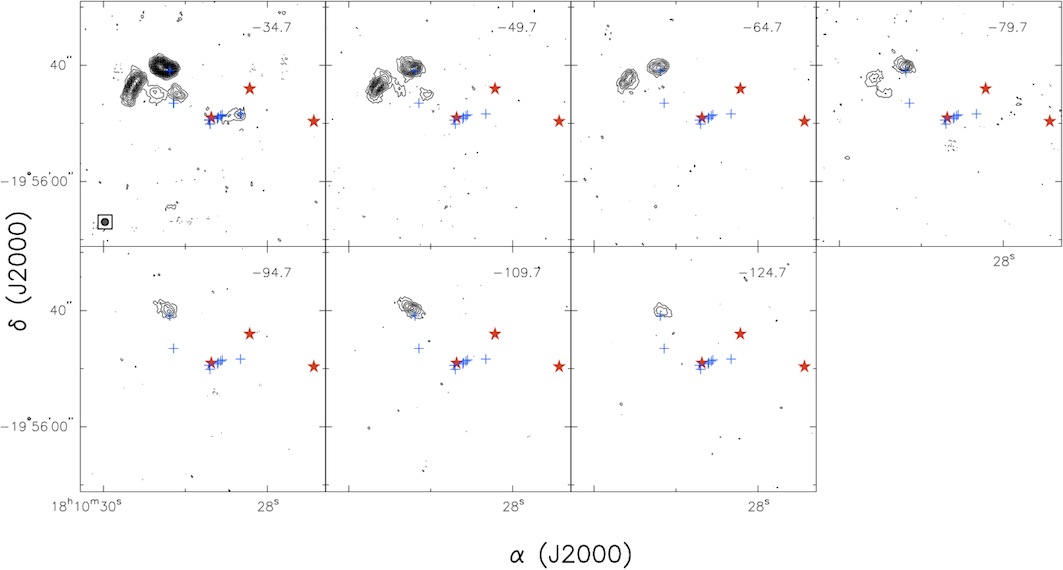}}
\caption{The channel maps of the $^{12}$CO (2-1) line. We bin per ten 1.5 kms$^{-1}$ velocity channels into one 15 kms$^{-1}$ velocity channel.
Solid contours start from 0.03 Jy/beam with 0.03 Jy/beam intervals; dashed contours start from -0.03 Jy/beam with -0.03 Jy/beam intervals. 
}
\label{fig_cochan4}
\end{figure}

In Figure \ref{fig_hcnpv}, we present the pv diagrams of HCN (3-2), SiO (8-7) and $^{12}$CO (2-1) around \textit{HCN outflow}, which are all cut in the RA direction.
From both panels in this figure, we can clearly see the broad outflow signatures around the angular offset of $\sim$-5$''$.
The HCN (3-2) emission and the $^{12}$CO emission consistently trace the outflow to the velocity of $\sim$-25 kms$^{-1}$.
Owing to the detection limit, the SiO pv diagram only traces the outflow in a limited velocity range.
From the pv diagram, we see that SiO is also excited in the hot toroid, which is located 2$''$--3$''$ east of the \textit{HCN outflow}.  
Assuming the LTE conditions, optically thin emission, and the excitation temperature of 100 K, we estimate the total number of the SiO molecule in \textit{HCN outflow} to be $\sim$1.5$\cdot$10$^{48}$.

We apply the same assumptions to derive the total number of the HCN and the $^{12}$CO molecules in the \textit{HCN outflow}. 
Adopting the galactic $^{12}$CO abundance (i.e. [$^{12}$CO]/[H$_{2}$]$=$1$\cdot$10$^{-4}$), we list the derived number and the implied abundance ratios in Table 4.
In the galactic molecular clouds, the  [HCN]/[$^{12}$CO] ratio ranges from 10$^{-4}$ in the quiescent zone to 2.5$\cdot$10$^{-3}$ in the hot core regions (Blake et al. 1987).
The enhanced [HCN]/[H$_{2}$] ratio has also been reported in other (proto--)stellar outflows (e.g. IRAS 20126: 0.1--0.2$\cdot$10$^{-7}$, Su et al. 2007; L1157: 5$\cdot$10$^{-7}$, J{\o}rgensen et al. 2004).
In the \textit{HCN outflow}, the value of [HCN]/[$^{12}$CO] in the blueshifted velocity range is consistent with an enhanced HCN abundance.
If our assumption of [$^{12}$CO]/[H$_{2}$] ratio is valid, the value of  [HCN]/[H$_{2}$] in the \textit{HCN outflow} is in between of the two referenced cases.
Even if the assumed [$^{12}$CO]/[H$_{2}$] ratio is overestimated by a factor of 10 (see the discussions in Section 3.2), the derived  [HCN]/[H$_{2}$] ratio in the  \textit{HCN outflow} is still higher than the reported typical value of $\le$7$\cdot$10$^{-9}$ (J{\o}rgensen et al. 2004).
The HCN molecule is usually regarded as a dense gas tracer.
Our results suggest, however, that the HCN emission can also be locally enhanced in outflows and shocks around the dense core, and therefore might not be a good tracer of the overall dynamics in massive cluster forming regions.

Among all detected high velocity outflows, the \textit{HCN outflow} does not have specifically higher linear momentum or energy.
The uniqueness in the excitation of SiO (8-7) may be explained by the heating of the central OB cluster, or be explained by the molecular gas erupted from the hot toroid with shock enhanced SiO abundance. 
Some hints can be seen from the 3.6 cm continuum image (Figure \ref{fig_hcn}, bottom panel), which shows an elongated emission feature around the location of the \textit{HCN outflow}.
Whether the elongated 3.6 cm emission feature is physically associated with \textit{HCN outflow} or is just a projection effect can be examined by future ALMA observations of molecular lines and the hydrogen recombination lines. 

We note that a few more elongated/filamentary features\footnote{Not the arc shaped structures isolated from the free--free emission peak.} are resolved in the previous 1.3 cm free--free continuum observation (Sollins et al. 2005).
Those elongated/filamentary features have the lengths of $\sim$1$''$ (0.03 pc), and width of $\le$0.1$''$ (0.003 pc). 
The physical properties and the formation mechanism of those elongated features are still uncertain.
Suppose those elongated/filamentary features of 1.3  cm free--free continuum emission are ionized gas which are undergoing thermal diffusion with velocity of 10 kms$^{-1}$, the $\le$0.003 pc widths imply that their age is no more than 300 years. 
If all those elongated/filamentary features can also be explained by ionized outflows, the velocities of the outflows can be estimated by
\[
\frac{\mbox{length}}{\mbox{age}} = \frac{0.03\mbox{ pc}}{300\mbox{ years}} \sim 100 \mbox{ kms$^{-1}$}.
\]
This outflow velocity is consistent with the measurement in hydrogen recombination line (line--of--sight velocity 60 kms$^{-1}$, Keto \& Wood 2006), although the ionized outflow is not explicitly resolved in the image.
The higher brightness of those elongated/filamentary features than the ambient regions may suggest that those outflows are originally ejected in molecular form with a much higher density, and then ionized by the stellar radiation. 
The molecular counterparts of those ionized outflows are not necessarily detected, especially in the bipolar direction, if stellar ionization is important. 
We suggest that the central OB cluster may accrete the molecular gas of very compact/clumpy structures (see the NH$_{3}$ opacity results, Sollins \& Ho 2005), which can survive the stellar ionization. 
The molecular gas can then approach individual O--type stars to the distance of  $\sim$10 AU, and be centrifugally accelerated to the outflow velocity.
Small scale accretion disks may exist around the O--type stars. 
However, they are not detectable with the current instruments owing to the beam dilution. 

\section{Discussions}
\label{chap_discuss}
In G10.6-0.4, previously we estimated the scalar momentum feedback from the stellar wind and from the ionized gas pressure to be of 10$^{2}$--10$^{3}$ M$_{\odot}$kms$^{-1}$ (Liu et al. 2010 b).
From the $^{12}$CO data presented in this paper, we conclude that the total momentum feedback from the protostellar outflows has the same order of magnitude.
However, these feedback mechanisms can still play very different roles in the massive cluster forming regions, owing to their differences in physical properties.

The stellar wind is an isotropic feedback mechanism.
By comparing the resulting star formation efficiency in numerical hydrodynamical simulations, Nakamura \& Li (2007) suggested that the \textit{spherical wind} cannot propagate efficiently in dense gas.
The feedback from such mechanisms is trapped in small size scale, where the induced turbulence is quickly dissipated. 
Therefore the energy feedback from the \textit{spherical wind} supports the cloud less efficiently.
The protostellar outflow is ejected from objects embedded in the local overdensities. 
Its momentum, however, is typically collimated in a small solid angle, which makes it penetrate the dense gas easily and can produce large scale disturbance.
The induced protostellar turbulence potentially replenishes the dissipated initial turbulence in large scale, and plays the role of regulating the cloud contraction and the star formation efficiency.  

From the numerical simulation, Mac Low (1999) suggested the characteristic timescale of turbulence dissipation $t_{d}$ and the free--fall timescale $t_{ff}$ have the relation:
\begin{equation}
t_{d} = \left(\frac{3.9\lambda_{d}}{M_{rms}\lambda_{J}}\right)t_{ff},
\end{equation}
where $\lambda_{d}$ is the driving scale of the turbulence, $\lambda_{J}$ is the Jeans length, and $M_{rms}$ is the rms Mach number of the turbulence.
Assuming a total mass of 1000--2000 M$_{\odot}$ is enclosed in the 0.5 pc scale envelope, the global free--fall timescale is about 10$^{5}$ years.
If the averaged gas temperature is 30--50 K, the Jeans length is about 0.1 pc.
The typical value of $M_{rms}$ can be 5--10\footnote{The scale of the initial turbulence in massive cluster forming regions can be referenced to Wang et al. 2008. Our present paper together with the previous observations on infrared dark cloud (Wang et al. 2008; Zhang et al. 2009) are consistent with a scenario of self regulated structure formation via turbulence dissipation and the outflow feedback.}.
Assuming the main driving source of the turbulence is the compact (proto--)stellar outflows, the value of $\lambda_{d}$ can be estimated by the physical size scale of the observed outflow systems, which is about 0.06 pc (2$''$).
Therefore, the characteristic timescale $t_{d}$ in G10.6-0.4 is about 0.23--0.47 times of t$_{ff}$, which is about a few times of 10$^{4}$ years.
Among the outflow systems detected in G10.6-0.4, the maximum outflow dynamical timescale is about 10$^{4}$ years, which is smaller than $t_{d}$.
The estimations of outflow energy in Section \ref{chap_estimation} suggests that the  energy injection from the protostellar turbulence is capable of balancing the turbulence energy dissipation in the relevant timescale.

We provide rough estimations for the total protostellar mass from a fiducial momentum ejection efficiency I$_{total}$ = P$_{*}$M$_{*}$, where I$_{total}$ is the total ejected scalar momentum, M$_{*}$ is the total stellar mass, and P$_{*}$ is a proportional factor.
Following Nakamura \& Li (2007), we adopt P$_{*}$= 50 kms$^{-1}$.
In G10.6-0.4, the projected scalar momentum in the blueshifted and the redshifted velocity ranges is 273 M$_{\odot}$kms$^{-1}$, which leads to $f_{los}$$\cdot$M$_{*}$ = 5.45 M$_{\odot}$, where $f_{los}$ is the ratio of the scalar momentum contributed by the line--of--sight velocity component to the total scalar momentum.
Statistically the value of $f_{los}$ can be estimated by \[\frac{\int_{0}^{\frac{\pi}{2}} \sin\theta d\theta d\phi}{\int_{0}^{\frac{\pi}{2}} 1 d\theta d\phi} \simeq 0.637, \]
assuming the inclination angle of the outflows are uniformly randomly distributed. 
Given the 1000--2000 M$_{\odot}$ total molecular mass in the envelope, the star formation efficiency (SFE) in the past 10$^{4}$ years is about 0.42\%--0.86\%.
Assuming a uniform star formation rate in the time domain, the star formation efficiency in one free--fall timescale is of the order of a few percent. 
This estimated SFE is comparable to the SFE in the simulations of Nakamura \& Li (2007), consistently suggesting the cloud contraction is self--regulated by the local star formation.
High resolution observations of the thermal dust emission to statistically study the prestellar and protostellar cores, will further improve the constraint on the efficiency of the outflow feedback.
The number of the YSOs can be statistically estimated by dividing M$_{*}$ with the averaged protostellar mass $\bar{m}$. 
The fiducial value of $\bar{m}$ based on the Scalo (1986) initial mass function (IMF) is 0.5 M$_{\odot}$.
In G10.6-0.4, the value of $\bar{m}$ can be biased by the evolutionary stage, and potentially the observational selection bias that we only pick up the systems with powerful molecular outflows.
Assuming the detected outflows are associated with YSOs, which on average have $\sim$1 M$_{\odot}$ stellar mass, the number of the YSOs can be estimated by M$_{*}$/1.0 = 8.56.
The surface density of these YSOs in the $\sim$0.5 pc region is therefore $\sim$8.56/(0.5$^{2}$) $\sim$ 34 pc$^{-2}$.
The derived YSO surface density should be treated as a lower limit since our observations do not trace protostellar objects which have weak or no outflows.

Alternatively, we can assume each of the compact high velocity outflows (R$_{1\cdots4}$, B$_{1\cdots6}$, $HCN$) is associated with one protostellar objects. 
Considering also the non--identified outflow systems in the intermediate velocity range, there can be $\sim$10 protostellar objects in total in the observed region.
If estimating the averaged mass of each protostellar object by 1 M$_{\odot}$, our observational results suggest that an order of $\sim$10 M$_{\odot}$ molecular gas are converted into (proto--)stellar objects in $\sim$10$^{4}$ years. 
The system potentially evolves into a stellar cluster with on order of 10$^{2}$ stars in a few t$_{ff}$.

The ionized gas only propagates in low density regions, which have low recombination rates.
However, our previous studies (Liu et al. 2010 b) suggest that the thermal pressure of the ionized gas is capable of driving the large scale (a few times of 0.1 pc) coherent motions, such as the expansional motions of the ionized cavities or the bubble walls.
The absolute velocity of such kind of motion is about a few kms$^{-1}$, and which is smaller than the thermal sound speed of the ionized gas ($\sim$10 kms$^{-1}$).
The efficiency of converting the kinetic energy in such expansional motion to the large scale turbulence energy has to be investigated in theoretical studies.

\section{Conclusion}
\label{chap_conclusion}
We present the arcsecond resolution interferometry observations of the $^{12}$CO (2-1) transition.
From the distribution and the mass/momentum/energy of the detected high velocity outflows, we discuss the role of the outflow feedback in the massive cluster formation region.
Our main results are:
\begin{itemize}
\item We detect multiple high velocity outflow components in the blueshifted and the redshifted velocity ranges, respectively. The total molecular mass in these outflow systems are about 10 M$_{\odot}$, and the total momentum and energy budgets are of the order of 10$^{2}$ M$_{\odot}$kms$^{-1}$ and 10$^{47}$ erg, respectively. 
\item No high velocity bipolar molecular outflow parallel to the global rotational axis is directly detected around the central 0.1 pc hot toroid. However, we cannot rule out the possibility that it is photo--ionized. 
\item The $^{12}$CO outflows are interacting with the dense molecular envelope, which can be diagnosed by the CH$_{3}$OH, HCN, and the SiO shock signatures.  
\item From the distribution of the UC H\textsc{ii} regions, we know that multiple massive stars have been formed in the 0.5 pc scale massive envelope, and not only formed in the central 0.1 pc hot toroid. From the association of high velocity outflows with water maser sources, and with the interaction signatures detected in thermal molecular emission, we suggest that multiple protostellar objects with $\sim$10 M$_{\odot}$ in total may also have been formed within 10$^{4}$ years. The star formation efficiency over one global free--fall timescale is of the order of a few percent. 
\item We detect strongly enhanced HCN (3-2) emission in a outflow system near the hot toroid, and therefore suggest that HCN is not a good tracer for the global dynamics.  
\end{itemize}
We will follow up the polarization observation to provide thorough discussions of the energetic relation in this source. 
We emphasize that the discussions and conclusions in this paper are relevant to the cluster forming regions with size scale of a parsec or a fraction of a parsec, and with mass of thousands of M$_{\odot}$.
In less massive systems (with lower opacity), the radiative hydrodynamical simulations suggest that the heating by the stellar radiation efficiently suppress the star formation (Offner et al. 2009).


\acknowledgments
Baobab Liu thanks the SMA staff to support the observations.
{\it Facilities:} \facility{SMA}
\bibliographystyle{aa}

\appendix
\section{A. The Channel Maps of the  $^{12}$CO (2-1) and the HCN (3-2) Lines}
\label{chap_channel}
We present the channel maps of the $^{12}$CO (2-1) line and the HCN (3-2) lines in this section.
Figure \ref{fig_cochan1}, \ref{fig_cochan2}, \ref{fig_cochan3}, and \ref{fig_cochan4} show the $^{12}$CO (2-1) line in the velocity channels of -131.5 kms$^{-1}$-- 77 kms$^{-1}$.
Contour levels are adjusted according to the rms noise. 
We think the channel maps in the redshifted velocity channels (9.5 kms$^{-1}$--45.5 kms$^{-1}$) are contaminated by the contribution of the foreground molecular gas, and therefore show subparsec scale diffuse emission, which are not seen in the channel maps in the blueshifted velocity range ( -124.75-- -15.25 kms$^{-1}$). 
The absorption line of $^{12}$CO (2-1) is also detected at the location of the free-free continuum peak (ra:18:10:28.683 dec:-19:55:49.07), up to $v_{lsr}$$\sim$45 kms$^{-1}$.
The HI absorption experiment (Caswell et al. 1975) consistently shows the foreground absorption up to $v_{lsr}$$\sim$45 kms$^{-1}$.

\begin{figure}
\resizebox{!}{16cm}{
\includegraphics[scale=1.2]{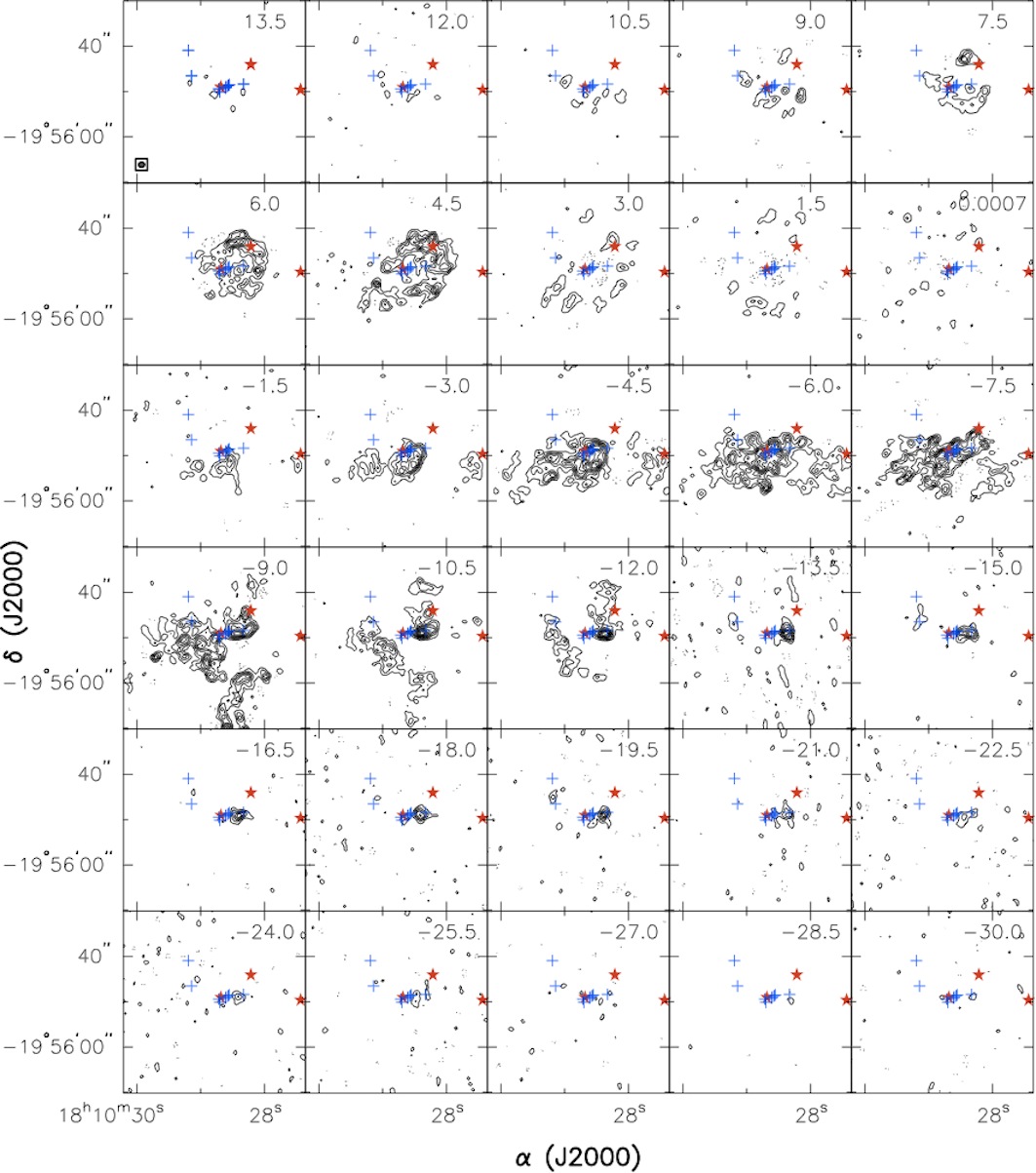}}
\caption{The channel maps of the HCN (3-2) line.  
Solid contours start from 0.55 Jy/beam with 0.55 Jy/beam intervals; dashed contours start from -0.55 Jy/beam with -0.55 Jy/beam intervals. 
Three 1.3 cm free-free continuum peaks (RA:18$^{h}$10$^{m}$28.683$^{s}$, Decl:-19$^{o}$55$'$49$''$.07; RA:18$^{h}$10$^{m}$28.215$^{s}$, Decl:-19$^{o}$55$'$44$''$.07 ; RA:18$^{h}$10$^{m}$27.435$^{s}$, Decl:-19$^{o}$55$'$44$''$.67) are marked by red stars. Cross symbols mark the water maser detections (Hofner \& Churchwell 1996).
The relative positional accuracy of the maser data is about 0.1$''$, which is much smaller than the size of the crosses.}
\label{fig_hcnchan1}
\end{figure}

Figure \ref{fig_hcnchan1} show the channel maps of the HCN (3-2) line.
We note the high consistency between the HCN and the $^{12}$CO channel maps.
In the HCN (3-2) channel maps, the low velocity counterpart of outflow R$_{3}$ (see Section \ref{chap_name}) may be visually identified in the velocity range of  7.5 kms$^{-1}$--12 kms$^{-1}$; and the low velocity counterparts of outflow B$_{3}$ and B$_{5}$ may be visually identified in the velocity range of  -10.5 kms$^{-1}$-- -15 kms$^{-1}$; the \textit{HCN outflow} can be seen in the same velocity range as covered by its $^{12}$CO emission

\section{B. The Position-Velocity Diagrams of the Interaction Signatures}
\label{chap_outflowpv}
\begin{figure}
\resizebox{!}{15cm}{
\includegraphics[scale=0.8]{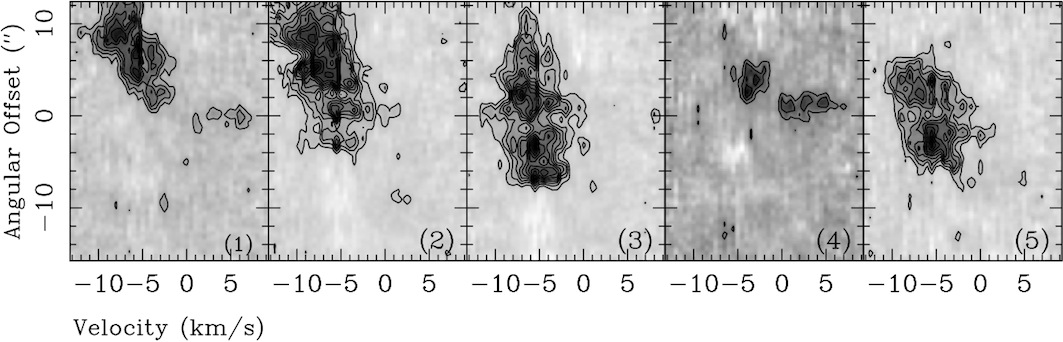}}
\caption{The pv diagrams of the CH$_{3}$OH 5(0,5)-4(0,4) A+ transition, at the locations of the broad CH$_{3}$OH interaction signatures (Section \ref{chap_broadch3oh}), and at the locations of two water maser sources. The centers of the pv cuts and their position angles are:
(1) RA=18$^{h}$10$^{m}$29.329$^{s}$, Decl=-19$^{o}$55$'$56$''$.79, PA=0$^{o}$;  (2) RA=18$^{h}$10$^{m}$29.088$^{s}$, Decl=-19$^{o}$55$'$53$''$.59, PA=0$^{o}$; 
(3) RA=18$^{h}$10$^{m}$28.975$^{s}$, Decl=-19$^{o}$55$'$49$''$.79, PA=0$^{o}$;  (4) RA=18$^{h}$10$^{m}$29.192$^{s}$, Decl=-19$^{o}$55$'$40$''$.93, PA=45$^{o}$; 
(5) RA=18$^{h}$10$^{m}$29.144$^{s}$, Decl=-19$^{o}$55$'$46$''$.52, PA=45$^{o}$.
The former three panels are the broad CH$_{3}$OH interaction signatures, and the later two are the water maser sources.
Contours start from 0.18 Jy/beam with 0.18 Jy/beam intervals. Grey scales are adjusted to optimize the contrast of structures in each panel.}
\label{fig_ch3ohoutflow_pv}
\end{figure}

Figure \ref{fig_ch3ohoutflow_pv} shows the pv diagrams of the CH$_{3}$OH 5(0,5)-4(0,4) A+ transition, at the locations of three broad CH$_{3}$OH interaction signatures, and at the location of two water maser sources.
Those CH$_{3}$OH interaction signatures are visually identified from the $f_{36}$ maps (Section \ref{chap_broadch3oh}).
The position angles of the pv cuts are chosen purposely to detect the broad line emissions and to avoid the confusion with nearby interaction signatures.
These figures provides complementary information in the velocity space, which cannot be explicitly seen in the $f_{36}$ maps.

The first three panels in Figure \ref{fig_ch3ohoutflow_pv} present the pv diagrams at three broad CH$_{3}$OH interaction signatures.
From these panels, we see broad emissions features of 1$''$--2$''$ angular sizescale around the zero angular offset, 
The last two panels in Figure \ref{fig_ch3ohoutflow_pv} present the pv diagrams at two water maser sources, which are associated with high velocity $^{12}$CO outflows (Section \ref{chap_name}). 
In panel (4), the broad emission signature can be seen  around the angular offset of 0$''$--2$''$; in panel (5), the broad emission signature can be seen around the angular offset of 1$''$--3$''$.

The angular offset of 3$''$ corresponds to the physical size scale of 0.09 pc.
Suppose the water maser sources mark the exact location of the stellar object, the 0.09 pc separation of the interaction signature from the water mater source is consistent with the typical size scale of the molecular outflow.
\clearpage

\begin{table}
\hspace{-1cm}
\begin{tabular}{lrccccc}
Transition & Frequency (GHz) &   E$_{u}$/k (K) & Einstein A--Coefficient  (s$^{-1}$)                            \\\hline\hline
CO (2-1)                               & 230.538      &  16.68  &  6.91$\cdot$10$^{-7}$  \\\hline

HCN (3-2)                             & 265.886      &  25.33  & 8.36$\cdot$10$^{-4}$   \\\hline

CH$_{3}$OH 5(0,5)-4(0,4) E   & 241.700      &  47.68  & 6.04$\cdot$10$^{-5}$    \\\hline

CH$_{3}$OH 5(0,5)-4(0,4) A+ & 241.791     &  34.65  & 6.05$\cdot$10$^{-5}$     \\\hline

CH$_{3}$OH 5(-2,4)-4(-2,3) E & 241.904      &  60.38  & 5.09$\cdot$10$^{-5}$     \\\hline
 
CH$_{3}$OH 5(2,3)-4(2,2) E  & 241.905      &  57.27  & 5.03$\cdot$10$^{-5}$     \\

CH$_{3}$OH 5(-3,3)-4(-3,2) E & 241.852      &  96.93  & 3.89$\cdot$10$^{-5}$      \\\hline

SiO (8-7)                              & 347.331      &  74.62  & 2.20$\cdot$10$^{-3}$      \\\hline      

\end{tabular}
\caption{Table of the selected molecular transitions. The quantum number of the transitions are listed in the first column. Their frequencies and upper-level-energy(E$_{u}$) are listed in the second and the third column. The Einstein A--Coefficient of each transition is listed in the fourth column. 
}
\label{table_molecule_list}
\end{table}

\clearpage

\begin{table}
\hspace{-1cm}
\scriptsize{
\begin{tabular}{lcccrc}
Transition           &   Beam                              & Channel width (kms$^{-1}$)    &   rms (Jy/beam)   &  SMA array configuration         & Observed Date \\\hline\hline
CO (2-1)            &   1$''$.2$\times$1$''$.2   &                                  1.5    &                0.03   & subcompact + compact + vex & 2009.02.09/2009.06.10/2009.07.12 \\

CH$_{3}$OH J=5 &   1$''$.5$\times$1$''$.3   &                                  0.53  &                0.06   & compact + vex                      & 2009.06.10/2009.07.12 \\

HCN (3-2)           &   1$''$.3$\times$1$''$.1   &                                 0.46   &                0.24   & compact + extended              & 2005.06.21/2005.09.09 \\

SiO (8-7)             &   4$''$.9$\times$1$''$.5   &                                  0.7    &                0.24   & compact                               & 2008.10.04 \\\hline

\end{tabular}
}
\caption{Table of the observational parameters.}
\label{table_parameters}
\end{table}

\clearpage

\begin{table}
\hspace{-2.3cm}
\scriptsize{
\begin{tabular}{ccccrccrr}
Outflow    &   ra  (J2000)   &   dec  (J2000)  &  Mass $M$ (M$_{\odot}$)   &  $I$$^{1}$ (M$_{\odot}$$\cdot$kms$^{-1}$)   & Energy $E$ (10$^{46}$ erg) & $t$ (yr) &  $\dot{I}$  (M$_{\odot}$$\cdot$kms$^{-1}$yr $^{-1}$) &  $\dot{E}$  (10$^{46}$ erg$^{-1}$yr$^{-1}$)  \\\hline\hline    
R1 &  18$^{h}$10$^{m}$29.464$^{s}$ & -19$^{o}$55$'$39.5$''$ &  0.18  &    3.2  &     0.12  & 4000 & 0.52$\times$10$^{-3}$ &  0.77$\times$10$^{-5}$\\                
R2 &  18$^{h}$10$^{m}$29.222$^{s}$ & -19$^{o}$55$'$42.3$''$ &  0.32  &    5.7  &     0.15  & 4100 & 1.39$\times$10$^{-3}$ &  3.77$\times$10$^{-5}$\\                
R3 &  18$^{h}$10$^{m}$29.953$^{s}$ & -19$^{o}$55$'$48.1$''$ &  0.56  &    8.1  &     0.15  & 6000 & 1.39$\times$10$^{-3}$ &  2.54$\times$10$^{-5}$\\                
R4 &   18$^{h}$10$^{m}$28.981$^{s}$ & -19$^{o}$56$'$02.1$''$ & 0.30  &    4.4  &     0.08  & 5300 & 0.85$\times$10$^{-3}$ &  1.46$\times$10$^{-5}$\\\hline        

B1 &   18$^{h}$10$^{m}$29.577$^{s}$ & -19$^{o}$55$'$42.7$''$ & 1.00  &  31.8  &     1.23  & 2800 & 11.55$\times$10$^{-3}$ &   38.50$\times$10$^{-5}$\\               
B2 &   18$^{h}$10$^{m}$29.194$^{s}$ & -19$^{o}$55$'$40.7$''$ & 1.82  &  57.2  &     2.62  & 2200 & 26.18$\times$10$^{-3}$ & 115.50$\times$10$^{-5}$ \\               
B3 &   18$^{h}$10$^{m}$29.095$^{s}$ & -19$^{o}$55$'$44.7$''$ & 0.40  &  7.93  &     0.15  & 2400 &   3.31$\times$10$^{-3}$ &     6.39$\times$10$^{-5}$\\               
B4 &   18$^{h}$10$^{m}$29.180$^{s}$ & -19$^{o}$56$'$04.7$''$ & 0.55  &  10.0  &     0.23  & 3300 &   3.00$\times$10$^{-3}$ &     7.01$\times$10$^{-5}$\\               
B5 &   18$^{h}$10$^{m}$29.322$^{s}$ & -19$^{o}$55$'$45.9$''$ & 0.18  &    3.6  &     0.08  & 2900 &   1.23$\times$10$^{-3}$ &     2.62$\times$10$^{-5}$\\               
B6 &   18$^{h}$10$^{m}$28.414$^{s}$ & -19$^{o}$55$'$43.9$''$ & 0.79  &  12.3  &     0.23  & 9500 &   1.31$\times$10$^{-3}$ &     2.46$\times$10$^{-5}$\\\hline       

HCN--B &  18$^{h}$10$^{m}$28.400$^{s}$ & -19$^{o}$55$'$49.1$''$ & 0.77  & 13.8 &  0.23  & 3300 &   4.16$\times$10$^{-3}$ &    7.01$\times$10$^{-5}$\\                  
HCN     &  18$^{h}$10$^{m}$28.400$^{s}$ & -19$^{o}$55$'$49.1$''$ & 7.53  & 55.5 &  0.62  & 8100 &   6.85$\times$10$^{-3}$ &    7.62$\times$10$^{-5}$\\\hline           
\end{tabular}
} 
\caption{A summary of the parameters of individual outflow systems. For R$_{i}$ outflows, the parameters are estimated in the  velocity range of 8.75-- 46.25 kms$^{-1}$; for B$_{i}$ outflows, the parameters are estimated in the velocity range of -124.75-- -15.25 kms$^{-1}$. The parameters of $HCN$--B outflow is estimated in the velocity range of -124.75-- -15.25 kms$^{-1}$; and the parameters of \textit{HCN outflow} is estimated in the same region as $HCN$--B outflow but for the entire velocity range (-124.75 kms$^{-1}$--58.25 kms$^{-1}$).
The outflow dynamical timescale $t$ is estimated by dividing the projected size scale with the averaged line--of--sight velocity $I$/$M$.
\footnotesize{$^{1}$The line--of--sight momentum.}
}
\label{table_outflow}
\end{table}

\begin{table}[h]
\begin{center}
\begin{tabular}{crcccc}
Velocity Range      &                                     & n$_{HCN}$                 &  n$_{^{12}CO}$               &   [HCN]/[$^{12}$CO]    &  [HCN]/[H$_{2}$]  \\\hline\hline
Blue Shifted & (-124.75-- -15.25 kms$^{-1}$)  & 2.4$\cdot$10$^{49}$  & 3.0$\cdot$10$^{52}$     &   0.8$\cdot$10$^{-3}$ &   0.8$\cdot$10$^{-7}$ \\
All  & (-124.75--58.25 kms$^{-1}$)             & 1.3$\cdot$10$^{50}$  & 2.9$\cdot$10$^{53}$    &   0.4$\cdot$10$^{-3}$ &   0.4$\cdot$10$^{-7}$ \\\hline
\end{tabular}
\caption{The derived HCN abundance ratio in the \textit{HCN outflow} (Section 3.3.2). The derived total numbers of HCN and $^{12}$CO for entire velocity range is uncertain, because in the intermediate velocity range, the optically thin assumption may not be valid, and the missing flux and the high optical depth of the foreground gas lead to the uncertainties in flux measurements.}
\end{center}
\label{table_hcn}
\end{table}
\clearpage

\end{document}